\documentclass[twocolumn,showpacs,floatfix,
superscriptaddress,
longbibliography,
amsmath,amssymb,
aps,
prb]{revtex4-2}

\usepackage{graphicx}% Include figure files
\usepackage{dcolumn}% Align table columns on decimal point
\usepackage{amsmath}
\usepackage{bm}% bold math
\usepackage{appendix}
\usepackage{soul}
\usepackage{multirow}
\usepackage[bookmarks=true,colorlinks=true,urlcolor=blue,linkcolor=blue,citecolor=blue,breaklinks]{hyperref}
\usepackage[mathlines]{lineno}% Enable numbering of text and display math
\usepackage{inconsolata} % very nice fixed-width font included with texlive-full
\usepackage[usenames,dvipsnames]{color} % more flexible names for syntax highlighting colors
\usepackage{listings}
\usepackage{xcolor}
\usepackage{bbm}
\usepackage{braket}
\usepackage{pdfpages} % include pdfs
\usepackage{pgffor} % for loops
\usepackage{xr} % referencing the supplement
\usepackage[normalem]{ulem}

\definecolor{AO}{rgb}{0.0, 0.5, 0.0}   % darkgreen

\makeatletter
\AtBeginDocument{\let\LS@rot\@undefined}
\makeatother

\renewcommand{\selectlanguage}[1]{}
\begin{document}

\preprint{APS/123-QED}

\title{Symmetry-Induced Logarithmic Relaxation in the Quantum Kicked Rotor}

\author{Julien Hébraud}
\affiliation{Laboratoire de Physique Théorique, Université de Toulouse, CNRS, UPS, France.}

\author{Floriane Arrouas}
\affiliation{Laboratoire Collisions Agrégats Réactivité, Université de Toulouse, CNRS, 31062 Toulouse, France}

\author{Bruno Peaudecerf}
\affiliation{Laboratoire Collisions Agrégats Réactivité, Université de Toulouse, CNRS, 31062 Toulouse, France}

\author{Juliette Billy}
\affiliation{Laboratoire Collisions Agrégats Réactivité, Université de Toulouse, CNRS, 31062 Toulouse, France}
\author{David Guéry-Odelin}
\affiliation{Laboratoire Collisions Agrégats Réactivité, Université de Toulouse, CNRS, 31062 Toulouse, France}

\author{Olivier Giraud}
\affiliation{Université Paris-Saclay, CNRS, LPTMS, 91405, Orsay, France.}

\author{Bertrand Georgeot}
\affiliation{Laboratoire de Physique Théorique, Université de Toulouse, CNRS, UPS, France.}

\author{Gabriel Lemarié}%
\email{ gabriel.lemarie@cnrs.fr}
\affiliation{Université Côte d’Azur, CNRS, INPHYNI, Nice, France}

\author{Christian Miniatura}
\email{christian.miniatura@cnrs.fr}
\affiliation{Université Côte d’Azur, CNRS, INPHYNI, Nice, France}
\affiliation{Centre for Quantum Technologies, National University of Singapore, Singapore.}

\date{\today}% It is always \today, today,

\begin{abstract}
We study the effect of discrete symmetries on coherent multiple scattering in the quantum kicked rotor. When the initial momentum is set to zero — as in recent Bose–Einstein condensate experiments— the effective pseudo-disorder becomes even under momentum inversion. The resulting discrete mirror symmetry of the dynamics profoundly alters spectral correlations: it generates quasi-degenerate Floquet doublets localised at opposite momenta, whose exponentially small splittings produce a hierarchy of exponentially large dynamical timescales. The coherent backscattering and forward-scattering peaks then exhibit a striking non-monotonic evolution and strongly asymmetric contrasts, followed by an exceptionally slow logarithmic relaxation toward a common asymptotic value — a hallmark of glassy dynamics, here emerging in a fully coherent quantum system. That such archetypal glass-like behaviour arises from a single discrete symmetry constraint reveals an unexpected and deep connection between quantum coherence and slow relaxation phenomena.
\end{abstract}

\maketitle

\section{Introduction} 

Symmetries play a central role in the physics of disordered or chaotic quantum systems. The presence or absence of time-reversal, spin-rotation, or chiral symmetries determines the appropriate universality class and strongly constrains spectral correlations, transport properties, and localisation phenomena \cite{anderson1958,abrahams1979,evers2008,altland1996}. A well-known example is the symplectic class, where spin–orbit coupling leads to weak anti-localisation and allows a metal–insulator transition even in two dimensions, in contrast with the orthogonal class \cite{hikami1980,evers2008}. Beyond symmetry-class changes, additional discrete symmetry constraints can further structure the spectrum and affect the dynamics in more subtle ways, see e.g. \cite{PhysRevE.50.145, doggen2017chaos, arnal2020chaos}.

Quantum interference in disordered systems manifests also itself through coherent multiple scattering. In spatially disordered media, coherent backscattering (CBS) arises from constructive interference between counter-propagating paths and constitutes a hallmark of weak localisation \cite{akkermans2007}. More recently, coherent forward scattering (CFS) has been identified as a distinctive signature of Anderson localisation \cite{karpiuk2012}. In quasi-one-dimensional systems, the CBS and CFS peak contrasts evolve monotonically and saturate on the localisation timescale, reflecting universal spectral correlations \cite{karpiuk2012,evers2008}.
The quantum kicked rotor (QKR) provides a paradigmatic platform to investigate such interference effects in a controlled setting \cite{chirikov1979universal,Casati1979,haake1991quantum}. Although classically diffusive in momentum space, the QKR exhibits dynamical localisation due to quantum interference, with exponentially localised Floquet eigenstates. Because the free-evolution phases between kicks act as pseudo-random on-site energies, the QKR can be mapped onto a quasi-one-dimensional Anderson model \cite{grempel1984quantum}. This mapping has enabled experimental observations with atomic matter waves of Anderson localisation in 1D \cite{moore1995atom} and of the Anderson transition in 3D \cite{chabe2008experimental}.

In the QKR, localisation occurs in momentum space; consequently, CBS and CFS peaks appear in the conjugate spatial domain \cite{lemarie2017}. While CFS was predicted theoretically more than a decade ago \cite{karpiuk2012}, it has only recently been observed experimentally in two independent realizations of kicked-rotor physics: indirectly via the return probability \cite{hainaut2018controlling}, and very recently by direct measurement using a Bose–Einstein condensate in a shaken optical lattice \cite{arrouas2025coherent}. The latter experiment required preparing an extremely narrow initial wave packet in real space—much narrower than the spatial period of the lattice potential. This strong squeezing was achieved via optimal control techniques~\cite{arrouas2025coherent,Dupont2021}.

A key feature in such experiments is that the quasi-momentum $\beta$ is conserved during the dynamics. The narrow initial momentum distribution of the BEC about $\beta=0$ restricts the dynamics to an invariant subspace of the Hilbert space \cite{arrouas2025coherent}, within which the effective pseudo-disorder generated by the free-evolution phases becomes symmetric under momentum inversion. In other words, the initial state selects an evolution sector in which the disorder realisation is constrained by mirror symmetry. In the Anderson analogy, this corresponds to a tight-binding model with an even on-site potential under spatial inversion. 

The central question we address in this work is therefore: how does coherent multiple-scattering dynamics change when the effective disorder is constrained by such a parity symmetry? Importantly, the system remains in the orthogonal universality class; however, the additional discrete symmetry constraint qualitatively modifies spectral correlations across parity sectors.
We show that mirror symmetry enforces quasi-degenerate pairs of Floquet eigenstates localised at opposite momenta. These parity-induced doublets possess exponentially small quasi-energy splittings, which generate exponentially large dynamical timescales. As a consequence, the CFS peak exhibits a non-monotonic evolution featuring an exceptionally slow, logarithmic relaxation toward its asymptotic value. This behaviour contrasts sharply with the standard quasi-one-dimensional scenario, where CFS grows monotonically and saturates on the localisation timescale \cite{karpiuk2012}.

Logarithmically slow relaxation is typically associated with systems possessing complex energy landscapes or constrained dynamics, such as glasses and trap models \cite{bouchaud1992, bouchaud1998out, amir2012relaxations, PhysRevLett.92.066801, pollak2013electron, PhysRevB.88.085106, PhysRevLett.88.076101}. Here, however, it emerges in a fully coherent quantum system as a direct consequence of a discrete symmetry constraint on the dynamics. Crucially, this glassy-like relaxation is directly visible in coherent multiple-scattering observables, providing experimentally accessible probes of symmetry-induced spectral structures. The aim of the present work is to identify and characterise the microscopic mechanism responsible for this anomalous relaxation and to describe its manifestations across the localised, intermediate, and metallic regimes of the quantum kicked rotor.

\section{Overview}

We summarize in this section the state of the art, the motivations, and the main results of our work in a non-technical way. The following sections will describe the results in detail.

Localisation can be characterized through coherent multiple scattering, in particular via the CBS and CFS peaks \cite{akkermans2007,karpiuk2012,lee2014dynamics,PhysRevLett.112.110602,ghosh2014coherent,PhysRevLett.115.200602,PhysRevA.95.041602, lemarie2017,hainaut2018controlling,PhysRevB.97.041406,PhysRevResearch.3.L032044,10.21468/SciPostPhys.14.3.057,PhysRevResearch.6.L012021,arrouas2025coherent}, which emerge in the space conjugate to localisation. Let us summarize the essential results established for these interference features, first in (quasi-)one-dimensional spatial Anderson localisation with time-reversal symmetry (orthogonal universality class) and then in the corresponding case of dynamical localisation in momentum space for the standard QKR. In a spatially disordered system, one prepares an initial plane wave; after disorder averaging, the momentum distribution develops two interference peaks on top of a stationary incoherent background, reached at times larger than the mean free time. The coherent backscattering (CBS) peak appears at the momentum opposite to the initial one, and its contrast rapidly reaches unity while its width shrinks monotonically until the localisation time. Conversely, the coherent forward scattering (CFS) peak forms at the initial momentum; its contrast grows monotonically and reaches unity at the localisation time. At asymptotically long times, CBS and CFS become twin peaks of contrast~1 and width set by the inverse localisation length. These behaviours are fully understood in the quasi-one-dimensional orthogonal, unitary, and symplectic universality classes.

In the corresponding QKR in the orthogonal universality class, where dynamical localisation occurs in momentum space, these peaks instead appear in the spatial domain \cite{lemarie2017}. More precisely, one must start from an initial spatial distribution much smaller than the spatial period of the optical lattice used to kick the atoms. This requires the use of a BEC rather than a thermal cloud, as in the recent experiment of Ref.~\cite{arrouas2025coherent}. This choice can modify the correspondence with the (quasi-)1D spatial Anderson localisation discussed above. Indeed, the very narrow momentum distribution of the BEC, in practice peaked at zero momentum, constrains the effective pseudo-disorder in momentum to be even. In the Anderson analogy, this would correspond to an on-site potential which is symmetric under spatial inversion.

Our present analysis of coherent multiple-scattering signatures shows that such a discrete symmetry constraint fundamentally modifies the behaviour of interference peaks. Instead of the standard (quasi-)1D scenario in the orthogonal symmetry class—where CBS and CFS contrasts evolve monotonically toward their asymptotic values on the timescale of the localisation time—we find non-monotonic dynamics together with an exceptionally slow, logarithmic relaxation toward the long-time limit, as illustrated in Fig.~\ref{fig:3 regimes}. 
Furthermore, the asymptotic contrasts are strongly modified by the presence of the symmetries, a behaviour that was directly measured in \cite{arrouas2025coherent}. The asymptotic CBS and CFS contrasts were observed to be higher than 1 in the metallic regime, as confirmed by our numerical data shown in Fig.~\ref{fig:3 regimes} (c) and (f).

The remainder of the paper explains in detail the mechanisms responsible for this logarithmic relaxation to a modified contrast, focusing on the case where parity symmetry is present-which is related to the case of a spatially-symmetric disorder potential in the Anderson analogy.
In brief, the mirror symmetry generates quasi-degeneracies between pairs of states forming symmetric and antisymmetric combinations of wavefunctions localised at opposite momenta $p_0$ and $-p_0$, with $p_0$ ranging between the localisation length $\xi$ and the system size.
If these doublets were exactly degenerate, the CBS and CFS contrasts would reach $1$ and $3$, respectively, at times larger than the localisation time—a first nontrivial consequence of the symmetry. In reality, however, the degeneracy is only approximate and is lifted as time evolves. The energy splitting within each doublet is exponentially small, scaling as $\exp(-|p_0|/\xi)$, which generates an exponentially large associated timescale given by the inverse splitting. The ensemble of such exponentially large times for all doublets produces the observed logarithmically slow relaxation. Once the splittings of all relevant doublets have been resolved dynamically, both CBS and CFS contrasts eventually converge to their common long-time value of $2$.

The rest of the paper is organized as follows. In Section~III, we describe the theoretical model considered. Section~IV characterizes the symmetries of the model and their impact on Floquet eigenstates. Section~V provides a first qualitative description of the time evolution of the disorder-averaged position distribution. Section~VI addresses the CFS and CBS contrasts at time $t$, their relation to spectral form factors and the key role of the nearest-level spacing distribution and its emergent Shnirelman peak for the long-time logarithmic relaxation dynamics. In Section~VII, we discuss  the dynamical behaviours of the CFS and CBS peaks at different time scales, in particular the logarithmic relaxation at large times. Section~VIII concludes our Paper. Appendix~A discusses the CFS and CBS peak shapes while Appendix~B discusses some intermediate time dynamics aspects.

\section{Theoretical model} 

Our model is a variant of the long-known and well-studied periodically-kicked quantum rotor (QKR) \cite{Casati1979,moore1995atom,Raizen96,Ammann1998,Kanem2007,chabe2008experimental,Manai2015,Garreau2017}. The QKR is a paradigmatic model for quantum chaos displaying dynamical localisation, {\it aka} Anderson localisation in momentum space, see \cite{Santhanam22} for a recent review. Its (dimensionless) Hamiltonian is given by $\hat{H}(t) = \hat{p}^2/2 - K \cos\hat{x} \, \sum_{m\in\mathbbm{Z}} \delta (t-m)$, where $[\hat{x},\hat{p}] = i \hbar_e$. This model is easy to implement experimentally \cite{moore1995atom,Raizen96,chabe2008experimental,Manai2015,Cao2022,arrouas2025coherent} and its complex dynamics is solely controlled by two parameters, the dimensionless kick strength $K$ and the effective dimensionless Planck's constant $\hbar_e$. It describes a free-evolving particle that is periodically subjected to a cosine potential acting during a vanishingly short amount of time. Being time-periodic, the QKR model allows for a natural stroboscopic description of its evolution at integer times $t\in\mathbbm{N}$, $\ket{\psi_t} = U^t \ket{\psi_0}$. Here $U$, known as the quantum Floquet map, is the evolution operator over one time period. There is actually a continuous family of timing protocols, parametrized by the time at which the kick is applied, and for which all the corresponding Floquet maps are unitarily related \cite{thomas2025coherent}. The most-often used timing protocol is a full free evolution terminated by a kick, for which the corresponding Floquet map reads
\begin{equation}
    U_0 = e^{i \frac{K}{\hbar_e} \cos \hat{x}} \, e^{-i \frac{\hat{p}^2}{2 \hbar_e}}.
\label{eq:qkr0_evolution_operator}
\end{equation}
Writing the continuous momentum $p=(n+\beta)\hbar_e$, where $n\in\mathbbm{Z}$ and $\beta\in [-1/2,1/2[$ is the quasi-momentum, it is easy to show that successive kicks change $n$ but conserve $\beta$ since $\bra{n',\beta'}e^{i \frac{K}{\hbar_e} \cos \hat{x}}\ket{n,\beta} \propto \delta (\beta-\beta') \, J_{|n-n'|}(K/\hbar_e)$, where $J_{|n-n'|}$ is the Bessel function of integer order $|n-n'|$ \cite{Fishman82,grempel1984quantum}. This also shows that the kick-induced momentum change is restricted to a range of order $|n-n'| \sim \ell_K = K/\hbar_e$, which is identified as the scattering mean free path in momentum space.

Importantly, the QKR model does embody pseudo-randomness despite being deterministic. Indeed, when $\hbar_e$ is incommensurate with $\pi$, the free dynamical phases $\alpha (p) = p^2/(2\hbar_e) = \hbar_e (n+\beta)^2/2$ fill uniformly the unit circle and generate a pseudo-random sequence as a function of $n$ \cite{grempel1984quantum}. The QKR then maps directly onto a quasi-one-dimensional Anderson tight-binding model with pseudo-random on-site energies \cite{grempel1984quantum}. Moreover, in standard experiments employing a thermal cloud (from a magneto-optical trap), disorder averaging is implemented intrinsically: the initial atomic ensemble is well approximated by an incoherent mixture of plane waves with a continuous distribution of quasi-momenta $\beta$, so that observables naturally include an average over~$\beta$. From a theoretical point of view, a good-enough alternative is to consider the random QKR model (RQKR) where the dynamical phases are simply replaced by truly random phases $\alpha(p)$ uniformly distributed over $2\pi$, thus restoring the usual averaging over disorder configurations.  Introducing the Floquet eigenstates $\ket{\varphi_a}$ and quasi-energies $\omega_a$ solving 
\begin{equation}
    U\ket{\varphi_a} = e^{-i\omega_a} \ket{\varphi_a}
    \label{eq:Eigen}
\end{equation}
where $\omega_a \in [-\pi,\pi[$, the most important feature of these two quantum models is dynamical localisation \cite{Casati1979}: All the Floquet eigenstates are localised in the momentum $\mathbbm{Z}$-lattice; they are centred on the different momentum states and decay exponentially around them over a range called the localisation ``length" $\xi$. At large $K$, a theoretical estimate of $\xi$ is given by $\xi \sim \xi_K = (K/2\hbar_e)^2$ \cite{Shepelyansky1986,Fishman1989}. This implies that diffusion in momentum space, a hallmark of their corresponding classical versions, is stopped after the Heisenberg time $t_H \sim \xi$. 

To investigate the role of discrete symmetries in the time evolution of the quantum kicked rotor, it turns out to be more convenient (see below) to work with the ``half-kick" variant of the RQKR (HRQKR) introduced in \cite{thomas2025coherent} and characterized by the following Floquet map
\begin{equation}
   U= e^{i \frac{K}{2 \hbar_e} \cos \hat{x}} \, e^{-i \alpha(\hat{p})} \, e^{i \frac{K}{2 \hbar_e} \cos \hat{x}}.
\label{eq:qkr_halk_kick_evolution_operator}
\end{equation}
All our simulations, analyses and calculations in this paper have been done using this ``half-kick" model. Importantly, $U_0$ and $U= V^\dag U_0 V$ are unitarily related through $V=e^{i \frac{K}{2 \hbar_e} \cos \hat{x}}$. This means that both unitaries have same Floquet quasi-energy spectrum and that their Floquet eigenstates are simply related by $V$. Since $V$ is deterministic and diagonal in position space, the statistical properties of these Floquet eigenstates in {\it position} space are the same, an important feature for the following discussions.

\section{Discrete symmetries of our model}

\subsection{Parity and time reversal symmetries}

The discrete symmetries relevant to our study are parity $P$ and time reversal $T$. The former is a linear operator while the latter is an {\it anti-linear} operator \cite{Sigwarth2022}. We have 
\begin{eqnarray}
 \label{eq:Sym}
 &&P \,(\hat{x}, \hat{p}, t) \,P^{-1} = (-\hat{x}, -\hat{p}, t) \\ &&T \,(\hat{x},\hat{p}, t) \,T^{-1} = (\hat{x}, -\hat{p}, -t)\\
 &&PT \,(\hat{x}, \hat{p}, t)\, (PT)^{-1} = (-\hat{x}, \hat{p}, -t)
\end{eqnarray}
Note that  $T^2=\mathbbm{1}$ here since our KR model is spinless.
It is easy to see that
\begin{eqnarray}
    PUP^{-1} &=& e^{i \frac{K}{2 \hbar_e} \cos \hat{x}} \, e^{-i \alpha(-\hat{p})} \, e^{i \frac{K}{2 \hbar_e} \cos \hat{x}} \\
    TUT^{-1} &=& e^{-i \frac{K}{2 \hbar_e} \cos \hat{x}} \, e^{i \alpha(-\hat{p})} \, e^{-i \frac{K}{2 \hbar_e} \cos \hat{x}}.
\end{eqnarray}
For {\it symmetric} disorder $\alpha(\hat{p}) = \alpha(-\hat{p})$, we find $PUP^{-1} =U$ and $TUT^{-1}=U^\dag$, meaning that the random symmetric KR is both parity and time reversal invariant, which has strong consequences on its dynamics. We note that symmetric phase disorder is consistent with the original KR model described by $U_0$ in Eq.~\eqref{eq:qkr0_evolution_operator} where the free dynamical phases are indeed symmetric in momentum. There is however a subtlety to highlight here. As we have seen, the KR model conserves the quasi-momentum $\beta$. This means that the KR dynamics unfolds at {\it constant} $\beta$. Since $P$ or $T$ reverses $\beta$, this means that the original KR model is dynamically invariant under $P$ and $T$ {\it only when} $\beta=0$. 
To achieve this experimentally to a good approximation requires the use of an ultracold gas, initially prepared in the lowest energy state of the kicking cosine potential~\cite{arrouas2025coherent}. From a theoretical perspective, it simply amounts to working with $2\pi$-periodic wavefunctions in position space (in other words, the space variable becomes an angle) with momenta given by $p=n\hbar_e$ ($n \in \mathbbm{Z}$). In our numerical simulations, we have restricted momenta to the range $|n| \leq M \in \mathbbm{N}$ and imposed periodic boundary conditions, which means that $N=(2M+1)$ identifies with the lattice size and thus with the Hilbert space dimension.

\subsection{Symmetry properties of Floquet eigenstates}

\begin{figure}
    \centering
    \includegraphics[width=1\linewidth]{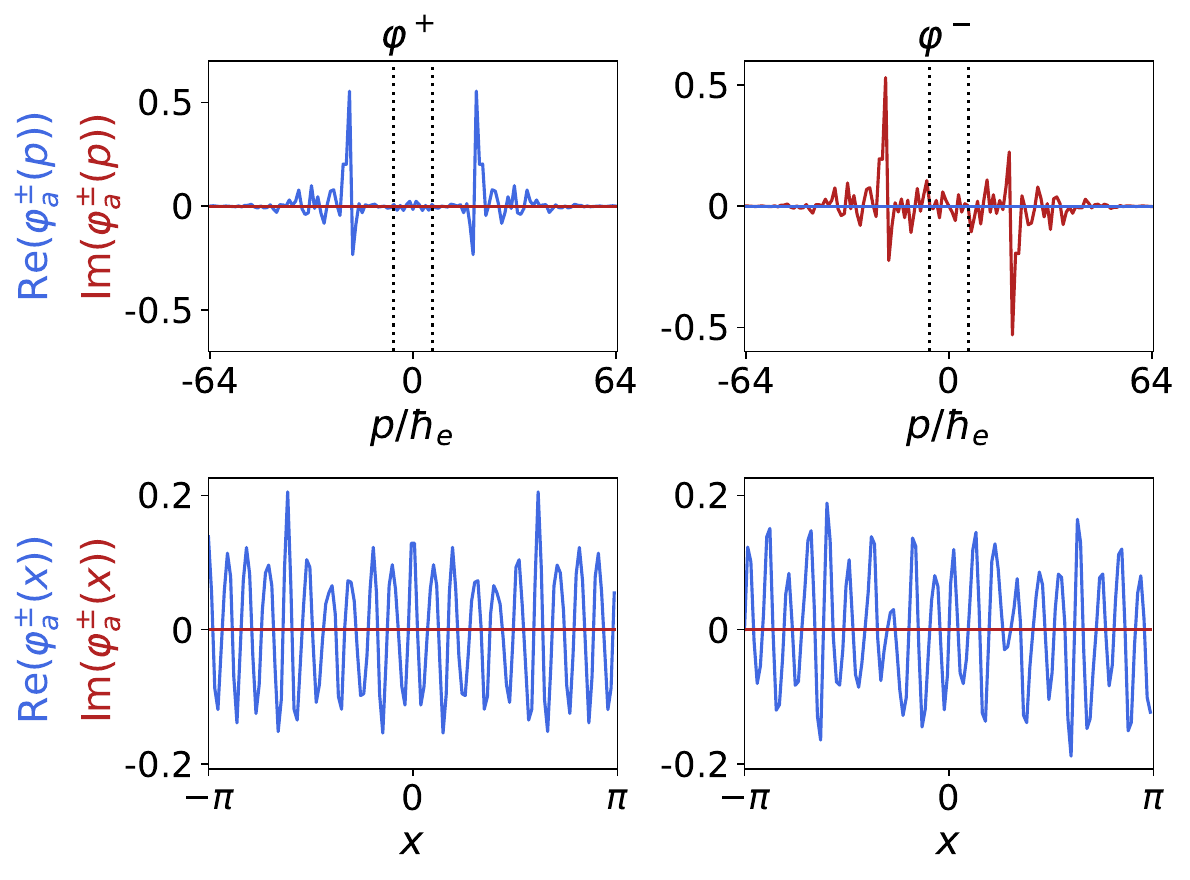}
    \caption{Two hybridized localised eigenstates in momentum space (top panels) and in position space (bottom panels) obtained for $N = 129$, $K = 10$ and $\hbar_e = 2$ ($\xi_K = 6.25$). We represent the real part (in blue) and the imaginary part (in red) of an even eigenstate $\varphi_a^+$ (left panels) and of an odd eigenstate $\varphi^-_a$ (right panels) lying outside
the central localisation box $|p| \leq \xi_K$ delimited by the vertical black dotted lines, but with the same localisation centres.
    The quasi-energy of the symmetric eigenstate $\varphi_a^+$ is $\epsilon_+ \approx -2.9548$ while that of the anti-symmetric eigenstate $\varphi_a^-$ is $\epsilon_- \approx -2.9568$, showing that these states form a doublet pair. As one can see, in position space both $\varphi_a^+$ and $\varphi_a^-$ are real and extended, while  in momentum space $\varphi_a^+$ is real and $\varphi_a^-$ is purely imaginary. These properties are a signature of the parity-induced hybridization mechanism similar to a quantum double-well that we discuss in Section \ref{sec:hybridization}.
    }
    
    \label{fig:eigenvectors}
\end{figure}

Assuming nondegenerate levels, the symmetry requirements $PUP^{-1}=U$ and $TUT^{-1}=U^\dag$ imply that the eigenstates of the Floquet map $U$ must satisfy $T\ket{\varphi_a} = \ket{\varphi_a}$ (such a choice of eigenstates is always possible since $T^2=\mathbbm{1}$ \cite{Sigwarth2022} and $P\ket{\varphi_a}=\pm \ket{\varphi_a}$). The full Hilbert space thus breaks into two parity sectors $\mathcal{H}=\mathcal{H}^{+}\oplus\mathcal{H^{-}}$ (one may note that $\dim\mathcal{H}^{+} = M+1$ while $\dim\mathcal{H}^{-} = M$). As a consequence, the Floquet spectrum of $U$ also breaks into two disjoint quasi-energy sets $\mathcal{S}^\pm$ associated to eigenstates $\ket{\varphi_a^{\pm}}$ in $\mathcal{H}^\pm$. 

Time-reversal invariance also implies $\varphi^*_a(x)=\varphi_a(x)$ and thus $\varphi_a(x)$ is {\it always real}. Parity-symmetry on the other hand implies that $\varphi_a^+(-x)=\varphi_a^+(x)$ is even in space for $\mathcal{H}^+$ while $\varphi_a^-(-x)=-\varphi_a^-(x)$ is odd for $\mathcal{H}^-$. As a consequence, by Fourier transform, $\varphi_a^+(p)$ is real and even in momentum for $\mathcal{H}^+$ while $\varphi_a^-(p)$ is purely imaginary and odd for $\mathcal{H}^-$. Figure~\ref{fig:eigenvectors} illustrates these properties for two hybridized Floquet eigenstates obtained in the localised regime $\xi \ll N$.

\subsection{Hybridization of Floquet eigenstates}
\label{sec:hybridization}

Assume first non-symmetric phase disorder $\alpha (-p) \neq \alpha (p)$. All Floquet eigenstates $\varphi_a(p) \equiv F_a(p-a\hbar_e)$ are localised in momentum space: They are labelled by their localisation centre $a \in [-M,M]$ with $F_a(p)$ extending over a range $\sim \xi\hbar_e$. Consider now the case where $|a| \gg \xi$. This means that the two eigenstates $\varphi_{\pm a}(p)$ live on two well-separated supports. They thus experience different and uncorrelated local phase disorders and we expect the functions $F_{a}$ and $F_{-a}$ to be different. In other words, their quasi-energies are unlikely to be close. Assume symmetric phase disorder now: The local phase disorder patches around $\pm a$ are mirror images of each other. We thus expect $F_{-a}(p) \approx F_{a}(-p)$ and the odd and even eigenstates in momentum space (as requested by the parity symmetry) are simply obtained by hybridizing these functions, $\varphi^{\pm}_{a}(p) \sim F_a(p-a\hbar_e) \pm F_{a}(-p+a\hbar_e)$. Fig.~\ref{fig:eigenvectors} gives a glimpse at this hybridization process.

This situation is reminiscent of the paradigmatic double-well phenomenon (see also \cite{doggen2017chaos}), an analogy which is recovered by assuming an infinitely high barrier at $p=0$, separating the full system into two exact copies, one for $p<0$, the other for $p>0$, with same  eigenvalues and mirrored localised eigenstates. By lowering that fictitious barrier at $p=0$, we let the two subsystems be coupled by quantum tunneling. The $p<0$ and $p>0$ eigenstates hybridize, lifting the eigenvalues degeneracies. Such pairs of hybridized eigenstates, of opposite parity, have very close quasi-energies for $|a| \gg \xi$. We call them \emph{doublets}. We expect that this quasi-energy splitting $\Delta_a$ scales with their exponentially-small eigenstate overlap in momentum space, which entails that
$\Delta_a \sim \Delta \, e^{-|a|/\xi}$, where $\Delta$ is an energy scale of the system used to quantify level spacings. In our system, the total disorder-averaged density of states $\nu(\omega) = \frac{1}{N} \, \overline{\sum_{a\in\mathcal{H}} \delta(\omega-\omega_a)}$ is constant over the range $[0,2\pi[$ \cite{lemarie2017}. As such, we can use the mean quasi-energy spacing $\Delta = 2\pi/N$ to quantify quasi-energy separations within the full spectrum comprising $N$ states. The corresponding time scale is $t_\Delta= 2\pi/\Delta= N$, that is, the system size.

The smallest doublet splitting is obtained for $|a|=M \sim N/2$, that is, for eigenstates localised near the boundaries of the system. As a consequence, these doublets can become exponentially close to each other in energy as the system size $N$ gets larger and larger. \\

\section{Disorder-averaged position distribution at time $t$}

\begin{figure*}
    \centering
    \includegraphics[width=\linewidth]{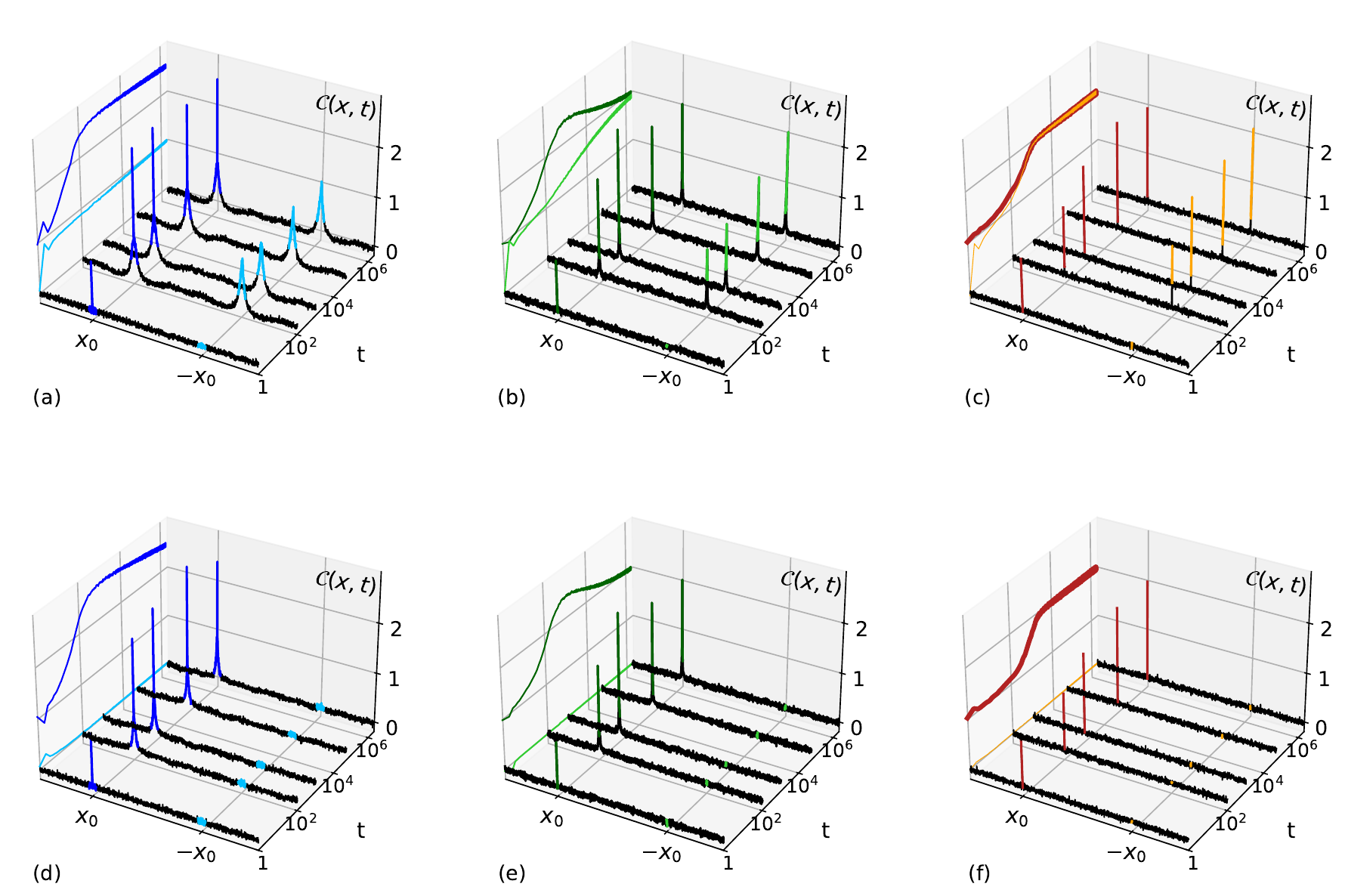}
    \caption{Time evolution of the contrast $C(x,t)$ for the Quantum Kicked Rotor when the system starts in a position state $\ket{x_0}$. Top panels (a-c): the kick potential is even and symmetric random phases are used (parity and time reversal symmetry are preserved). Bottom panels (d-f): the kick potential is no longer even but symmetric random phases are still used (parity is broken and time-reversal symmetry is preserved). Each panel shows $C(x,t)$ at times $t = 1, 101, 1001, 50 \hspace{2pt} 819$ and $1 \hspace{2pt} 999 \hspace{2pt} 998$ together with the time evolution of the CFS ($x=x_0$, dark colours) and CBS ($x=-x_0$, bright colours) peak values. For all plots, we have $M = 1024$ (lattice size $N=2M+1=2049$), $\hbar_e = 1$ and $x_0 = -\pi/2$. Panel (a): localised regime ($K = 6.32$, $\xi \sim \xi_K \approx 10 \ll M$). Panel (b): Intermediate regime ($K = 20$, $\xi \sim \xi_K \approx 100 < M$). Panel (c): Metallic regime ($K = 63.2$, $\xi \sim \xi_K \approx 1000 \approx M$). Panels (d,e,f): Same regimes when time-reversal is the sole symmetry of the system. Parameters are the same with the exception of (e) ($K = 11.2$). For readability, the CFS and CBS peaks have been coloured on the curves over a width equal to $2/\xi_K$, which is supposed to be the peak width at long times since $\xi \sim \xi_K$. 
    The symmetric random phases $\alpha(p)$ are chosen randomly over a range $2\pi$ for $p \geq 0$ and duplicated for $p<0$. Data are then averaged over 1080 realizations of disorder. The contrast curves are smoothened for readability for $t>2$, with a moving average of width 80.}
        \label{fig:3 regimes}
\end{figure*}

Starting the system in some given position state $\ket{x_0}$, we have $\langle x\ket{\psi(t)} = \psi(x,t)= \bra{x}U^t\ket{x_0}$ and the disorder-averaged position distribution reads $n(x,t) = \overline{|\psi(x,t)|^2}$, where $\overline{(\cdot\cdot\cdot)}$ denotes the average over phase disorder configurations. From the spectral decomposition \eqref{eq:Eigen} of $U$, it is easy to see that 
 \begin{equation}
    n(x,t) = \overline{\sum_{(a,b)\in \mathcal{H}} e^{-i\omega_{ab}t} \varphi_a(x_0)\varphi_a(x)\,\varphi_b(x_0)\varphi_b(x)},
    \label{eq:x0density}
\end{equation}
 where $\omega_{ab}=\omega_a-\omega_b$ encapsulates energy spacings.  
 
As shown in \cite{lemarie2017}, one can write $n(x,t)=n_{DB}(x)+n_I(x,t)$ where $n_{DB}(x)$ is the diffusive background obtained after classical isotropisation and where $n_I(x,t)$ is the interference contribution \footnote{Note that probability conservation imposes $\int_{-\pi}^{\pi} n(x,t) \, dx/(2\pi) =1$ and $\int_{-\pi}^{\pi} n_{DB}(x) \, dx/(2\pi) = 1$ so that $\int_{-\pi}^{\pi} n_I(x,t) \, dx/(2\pi) =0$.}. The interference contrast is then defined as $C(x,t)=n_I(x,t)/n_{DB}(x)$. Since $n_{DB}(x)=1/N$ for our KR model \cite{lemarie2017}, we have
\begin{equation}
    C(x,t) = Nn(x,t)-1.
\end{equation}
In Fig.~\ref{fig:3 regimes}(a-c), we show the time evolution of $C(x,t)$ for our $P$- and $T$-invariant HRQKR model (symmetric phase disorder). We observe two interference peaks located at $\pm x_0$ on top of an essentially flat  background. These peaks are the celebrated CBS (at $-x_0)$ and CFS (at $x_0$) peaks \cite{karpiuk2012, ghosh2014coherent, lemarie2017}. In the usual KR case where $PT$ is the sole symmetry of the system, these peaks become twin peaks in the long-time limit and achieve a contrast equal to 1 \cite{lemarie2017}. In the presence of symmetric phase disorder, the CBS and CFS peaks behave strikingly differently. In the localised regime $\xi \ll M$ (Fig.~\ref{fig:3 regimes}a), the CBS and CFS peaks grow and stabilize in time but their contrasts never equalize at the longest time explored: the CFS plateaus at a value well above 2 while CBS sticks to 1. In the intermediate regime $\xi \lesssim M$ (Fig.~\ref{fig:3 regimes}b), the CFS peak grows, its contrast overshoots the value 2 and then slowly returns back to 2 in the long-time limit. In the meantime, the CBS contrast grows monotonically to 2. Finally, in the metallic regime $\xi \gtrsim M$ (Fig.~\ref{fig:3 regimes}c), both contrasts grow monotonically and plateau at 2. One can also note, as is well visible in the localised regime, that the shapes of the CBS and CFS peaks are rather unusual: They show a delta-like spike on top of the usually observed bump with a finite width (see Appendix A). 

In Fig.~\ref{fig:3 regimes}(d-f), we show the contrast $C(x,t)$ obtained with the same parameters except for the use of a kick potential $V_T=- K \left(\cos\hat{x}+\tfrac{1}{2}\sin 2\hat{x}\right) \, \sum_{m\in\mathbbm{Z}} \delta (t-m)$ which is not position-symmetric. In this case, the $T$ symmetry is maintained but the $P$ symmetry is broken. The observed behaviours are qualitatively analogous to those observed in Fig.~\ref{fig:3 regimes}(a-c) with the notable absence of the CBS peak since the $PT$ symmetry is also broken \cite{lemarie2017}.

As we show in the rest of this paper, all these observations can be explained by the impact of symmetries on the quasi-energy spectrum and the Floquet eigenstates. 

\section{CFS and CBS contrasts at time $t$} 

\subsection{Contrasts asymmetry}

From Fig.~\ref{fig:3 regimes}, we see that the dynamics is essentially concentrated in the CFS and CBS peak contrasts, $C(x_0,t)$ and $C(-x_0,t)$ respectively. Sorting out even and odd eigenstates, we have
 $\psi(x,t) = \psi_{+}(x,t)+\psi_{-}(x,t)$, where 
 \begin{equation}
     \psi_\pm(x,t) =\sum_{a \in \mathcal{H}^\pm} e^{-i\omega_at} \varphi_a(x_0)\varphi_a(x)
 \end{equation}
is even/odd in space (from now on we omit the $\pm$ exponent on the Floquet states $\varphi_a^{\pm}$ since the information about the symmetry sector is contained in the label $a$). This allows us to decompose $n(x,t)=n_W(x,t)+n_B(x,t)$ into components within ($W$ subscript) and between ($B$ subscript) parity sectors 
 \begin{eqnarray}
 \label{eq:EOn}
     &&n_W(x,t)= \overline{|\psi_+(x,t)|^2} + \overline{|\psi_-(x,t)|^2}\\
     &&n_B(x,t) = 2 {\textrm Re}[\overline{\psi^*_-(x,t)\psi_+(x,t)}].
 \end{eqnarray}
 One should note that $n_W$ is even in $x$ while $n_B$ is odd in $x$. This means that the contrast admits the parity decomposition $C(x,t) = C_W(x,t)+C_B(x,t)$ where $C_W(x,t)=Nn_W(x,t)-1$ is even in $x$ while  $C_B(x,t)=Nn_B(x,t)$ is odd in $x$. From this, we immediately infer that the CFS and CBS contrasts $C(\pm x_0,t)$ are asymmetric since
\begin{equation}
    C(\pm x_0,t) = C_W(x_0,t) \pm C_B(x_0,t).
    \label{eq:CBSCFS}
\end{equation}
This asymmetry is well visible in Fig.~\ref{fig:3 regimes} in the localised and intermediate regimes. For large enough $N$, we also expect the parity sectors to have the same statistical properties so that $\overline{|\psi_+(x,t)|^2}= \overline{|\psi_-(x,t)|^2}$ and $n_W(x,t) = 2N \overline{|\psi_+(x,t)|^2} $. We thus get 
\begin{eqnarray}
\label{eq:Cplus} 
&&C_W(x_0,t) =
2N\ \overline{\sum_{\substack{a\in\mathcal{H}^+ \\ b\in\mathcal{H}^+}} e^{-i\omega_{ab}t}\varphi^{2}_a(x_0)\varphi^{2}_b(x_0)} - 1, \\
    &&C_B(x_0,t) = 2N\textrm{Re}\big[ \overline{\sum_{\substack{a\in\mathcal{H}^+ \\ b\in\mathcal{H}^-}} e^{-i\omega_{ab}t}\varphi^{2}_a(x_0)\varphi^{2}_b(x_0)}\big].
    \label{eq:Cminus}
\end{eqnarray} 
One may note in passing that $\psi_\pm(x_0,t) \equiv A_\pm(x_0,t)$ are nothing else than the parity-sorted position- and time-dependent spectral functions
\begin{equation}
    A_\pm(x_0,t) = \sum_{a\in\mathcal{H^\pm}} e^{-i\omega_a t} \, \varphi^2_a(x_0).
    \label{eq:SpectralF}
\end{equation}
At small enough times, doublets are not yet resolved and the $C_B(x_0,t)$ dynamics probes the usual level repulsion. We expect this dynamics to be essentially insensitive to parity aspects of the eigenstates. As such, we expect the double sums in Eqs.\eqref{eq:Cplus}-\eqref{eq:Cminus} to have similar time dependence at early-enough times. Since the double sum in Eq.\eqref{eq:Cplus} is nothing else than $\overline{|A_+(x_0,t)|^2}$, we deduce that, at small enough times, $C_B(x_0,t)$ is positive and grows in time, implying that the CFS peak is always higher than the CBS peak, as seen in Fig.~\ref{fig:3 regimes}. 

In the metallic regime, since the parity-related eigenstates can be distinguished, we also see that $C_{B}(x_0,t) \approx 0$ and the two peak contrasts adopt the same temporal behaviour dictated by $C_W(x_0,t)$ solely. For a finite system size $N$, all energy differences are resolved when $t\to\infty$ and only the diagonal terms $(a=b)$ survive in Eqs.~\eqref{eq:Cplus}-\eqref{eq:Cminus}. We thus have
\begin{eqnarray}
&&C_B(x_0,t\to\infty) = 0 \\
&&C_W(x_0, t\to\infty) = 2N\overline{\sum_{a \in \mathcal{H}^+} \varphi^4_a(x_0)} - 1.
\label{eq:ninfty}
\end{eqnarray}
As a consequence, the CBS and CFS peaks assume the same height once {\it all} quasi-energy differences are resolved. Furthermore, the random eigenstate amplitudes $\varphi_a(x)$ being real, their statistical properties within each parity sector can be inferred from the Porter-Thomas distribution with Dyson index $\beta=1$  \cite{porter1956fluctuations, falcao2022wave} and we have 
\begin{equation}
\sum_{a\in\mathcal{H}^{+}}\overline{\varphi^4_a(x_0)} = 3 \sum_{a\in\mathcal{H}^{+}}\Big(\overline{\varphi^2_a(x_0)}\Big)^2 
\label{mu4mu2}
\end{equation}
and $\overline{\varphi^2_a(x)} \sim 1/N$ for each parity sector in the limit $\ell\ll\xi\ll M$. We thus find
\begin{equation}
C(\pm x_0,t\to\infty) = C_\infty = 2,
\label{Cinfty}
\end{equation}
as seen in Fig.~\ref{fig:3 regimes}b, Fig.~\ref{fig:3 regimes}c where all energy differences are indeed resolved and the infinite-time plateau is reached. 

In the localised regime, as seen in Fig.~\ref{fig:3 regimes}a, the asymmetry of the CFS and CBS contrasts does not reduce to zero over the simulation time. This means that some energy differences have not been resolved. The case of Fig.~\ref{fig:3 regimes}a requires a little attention. Indeed, with the parameters chosen in this case, $\xi/\ell = K/(4\hbar_e) \approx 1.6$ is too small for the system to be in the universal regime. In this case, as explained in \cite{lee2014dynamics}, the eigenfunctions statistics deviates from its expected RMT and Porter-Thomas behaviours, see Appendix~B.

Obviously, $C_W(x_0,t)$ in Eq.~\eqref{eq:Cplus} involves time-oscillating terms where both quasi-energies $\omega_a$ and $\omega_b$ are drawn from the {\it same} statistical ensemble $\mathcal{S}^+$ (or $\mathcal{S}^-$). It is important to note that level spacings satisfying $\omega_{ab}t \ll 1$ are not resolved at time $t$ since $\exp(-i\omega_{ab}t) \approx 1$. This means that, as time increases, smaller and smaller level spacings are probed by the dynamics. As such, the evolution time scale associated to $n_{W}(x,t)$ should be the Heisenberg time $t_H\sim \xi$ as set by the usual level-repulsion properties within each statistical ensemble $\mathcal{S}^\pm$, that we expect to be identical for both ensembles. 

On the contrary, $C_B(x_0,t)$ in Eq.~\eqref{eq:Cminus} involves oscillatory terms where $\omega_a$ is drawn from $\mathcal{S}^+$ while $\omega_b$ is drawn from $\mathcal{S}^-$. As seen previously, when $\xi \ll M$, eigenstates localised outside the central box $[-\xi,\xi]$ come by even/odd doublet pairs with exponentially small splittings. This means that the long-time dynamics of $C_B(x_0,t)$ should probe these doublets. As time increases, they are progressively resolved until the last remaining one obtained when $|a|=M \sim N/2$. Introducing the number of localisation boxes in the system,
\begin{equation}
    N_L = \frac{N}{2\xi}
    \label{eq:NB},
\end{equation}
the time it takes to fully resolve the doublets is $t_D \sim e^{N_L} \, t_\Delta \sim N \, e^{N_L}$ and scales exponentially with $N_L$ in the localised regime. We have $t_D/t_H \sim 2N_L e^{N_L}$. Consequently, $t_D \gg t_H$ in the localised regime $N_L \gg 1$ and is just well beyond the evolution times explored in Fig.~\ref{fig:3 regimes}a. This is the reason why CBS and CFS heights reach different intermediate exponentially long-lived constant values, never reaching their equal infinite-time values in this case. 

In the metallic regime we have $N_L \lesssim 1$, and the support of eigenstates in momentum space always overlap. The hybridization phenomenon, still present, does not generically produce exponentially close quasi-energies. In this case, level repulsion is the rule and the CBS and CFS peaks have essentially the same dynamics. They reach their infinite-time plateau after the finite time $t_D \sim t_H$. 

Note that we also expect the short-time dynamics of $C_B(x_0,t)$ to unfold with the same time scale $t_H\sim \xi$ as $C_W(x_0,t)$ since similar level repulsion correlations are probed (i.e. this short time regime does not involve doublets). This means that in the metallic regime, we expect $C_B(x_0,t)$ to be bell-shaped with similar raising and decaying times.

As a final remark, we note that the hybridization mechanism leading to exponentially close energies also happens in the non-symmetric phase disorder case when the local disorder patches around $\pm p_a$ turn out to be similar. These occurrences, leading to doublet states, are exponentially-rare events since the parity symmetry is absent. Nevertheless, they do exist and drive the long-time behaviour of the CFS and CBS contrasts (``correlated volumes" approach \cite{mott1970conduction, altland1995spectral,ghosh2014coherent}). They also contribute to subtle transport mechanisms like (phonon-assisted) variable-range hopping \cite{mott1956transition, pollak2013electron}.

\subsection{Relation to spectral form factors} 

Let us now introduce the spectral form factor \cite{brezin1997spectral, altland2025statistics,akkermans2007}
\begin{equation}
    K(t) = \frac{1}{\dim \mathcal{H}} \, \overline{|\textrm{tr}U^t|^2} = \frac{1}{N} \, \overline{\sum_{(a,b)\in\mathcal{H}} \, e^{-i\omega_{ab}t}}.
    \label{eq:SFF}
\end{equation}
For our parity-symmetric system, we have 
\begin{equation}
    K(t)=K_W(t)+K_B(t),
\end{equation}
where
\begin{eqnarray}
&& K_W(t) = \frac{(M+1)K_{+}(t)+M K_{-}(t)}{N} \\
    && K_{\pm}(t) = \frac{1}{\dim \mathcal{H}^\pm} \ \overline{\sum_{(a,b) \in \mathcal{H}^{\pm2}} e^{-i\omega_{ab}t}}
    \label{eq:Kpmt}
\end{eqnarray}
and 
\begin{equation}
K_B(t) = \frac{2}{N} \ \overline{\textrm{Re}\big[\sum_{a\in\mathcal{H}^+,b\in\mathcal{H}^-} e^{-i\omega_{ab}t}\big]}.
\label{eq:KDt}
\end{equation}
As one can see, $K_{W}(t)$ encapsulate quasi-energy intra-correlations (i.e.~within each parity sector of the full Hilbert space) while $K_B(t)$ encapsulates quasi-energy inter-correlations (i.e. between the two parity sectors of the full Hilbert space). In the limit $N\gg 1$, the two parity ensembles have the same statistical properties and we have $K_{+}(t)=K_{-}(t)$, leading to 
\begin{equation}
K_W(t)= K_{+}(t).
\label{eq:FullK}
\end{equation}
The random eigenstates involved in the expression of $C_B(x_0,t)$, Eq.~\eqref{eq:Cminus}, belong to different parity subspaces. We can assume that they are uncorrelated in position space, 
so that $\overline{\varphi^2_a(x_0) \, \varphi^2_b(x_0)} \approx \overline{\varphi^2_a(x_0)} \ \overline{\varphi^2_b(x_0)} = 1/N^2$. Further assuming statistical independence between the random quasi-energy values and the corresponding random eigenstate amplitudes in space, Eq.~\eqref{eq:Cminus} yields
\begin{equation}
  C_B(x_0,t) = K_B(t).
  \label{eq:KDminus}
\end{equation}
At short enough times, we expect $K_+(t)$ and $K_B(t)$ to behave similarly, that is $K_B(t) \sim K^\infty_B \, K_+(t)$, where $K^\infty_B \leq 1$ is some proportionality factor depending on $N_L$.

Separating the diagonal and non-diagonal terms in the sum in Eq.~\eqref{eq:Cplus} for $C_W(x_0,t)$, and using Eqs.~\eqref{mu4mu2} and \eqref{Cinfty} for the diagonal part, we also get 
\begin{equation}
    C_W(x_0,t) = C_\infty + 2N\ \overline{\sum_{(a\neq b)\in\mathcal{H}^+} e^{-i\omega_{ab}t}\varphi^2_a(x_0)\varphi^2_b(x_0)}.
\end{equation}
To treat the remaining sum in the right-hand side, we call back the same statistical arguments used above for $C_B(x_0,t)$ to decorrelate the quasi-energies and the eigenstate amplitudes. All in all, since $C_\infty = 2$, we finally arrive at 
\begin{equation}
    C_W(x_0,t) = 1+ K_+(t).
\end{equation}
The contrast at the peaks, Eq.~\eqref{eq:CBSCFS}, then becomes
\begin{eqnarray}
    &&C(x_0,t)= 1 + K_+(t)+K_B(t) = 1 + K(t)
    \label{eq:C}\\
    &&C(-x_0,t) = 1 + K_+(t) - K_B(t).
\end{eqnarray}

\subsection{Relation to nearest-level spacing distributions}

In the right-hand side of \eqref{eq:SFF}, $t$ is not restricted to integer values; here we therefore consider that $t$ is a continuous variable. Introducing the reduced variables $\tau=t\Delta$, $s=\omega/\Delta$ and $s_{ab}= \omega_{ab}/\Delta$, where we recall $\Delta = 2\pi/N$, we have
\begin{eqnarray}
    &&K(\tau)= \frac{1}{N} \overline{\sum_{(a,b)\in \mathcal{H}} e^{-is_{ab}\tau}} \\
    &&K(s) = 2\pi \, \delta (s) + \frac{2\pi}{N}\overline{\sum_{(a\neq b)\in \mathcal{H}} \delta (s-s_{ab})},
    \label{eq:TFSFF}
\end{eqnarray}
where $K(s) = \int_{-\infty}^{\infty} d\tau \, K(\tau) \exp(-i s \tau)$ is the Fourier transform of $K(\tau)$. Following \cite{akkermans2007}, we introduce the probability $P(l,s)$ that two levels at a distance $s$ are separated by $l$ other levels. We thus have
\begin{equation}
    F(s) \equiv \sum_{l=0}^{N-2} \, P(l,s) = \frac{1}{N}\overline{\sum_{(a\neq b)\in \mathcal{H}} \delta (s-s_{ab})}.
    \label{twopointfunction}
\end{equation}
and $F(s)= F_W(s)+F_B(s)$, where $F_W(s)$ encapsulates spacings within each parity sector, while $F_B(s)$ encapsulates spacings between the parity sectors. This means that $F_W(s)$ controls $K_+(\tau)$, Eq.~\eqref{eq:Kpmt}, while $F_B(s)$ controls $K_B(\tau)$, Eq.~\eqref{eq:KDt}.  

The large-time convergence to $K(\tau \to\infty)=1$ is set by the behaviour of $F(s)$ when $s\to 0$. In this regime, $P(l+1,s) \ll P(l,s)$, such that $F(s\to 0) \approx P(0,s) \equiv P(s)$ reduces to the nearest-level spacing distribution and the long-time dynamics is dominated by short-ranged quasi-energy correlations. This means that $P(s) = P_W(s)+P_B(s)$ where $P_W(s)$ covers intra-parity nearest-level correlations and $P_B(s)$ covers inter-parity nearest-level correlations. It is important to note that both $P_W$ and $P_B$ cover the usual RMT level-repulsion aspects {\it and} very short-ranged correlations resulting from hybridization. Indeed, these exponentially small hybridization correlations do exist within $\mathcal{H}^\pm$ and are responsible for the long-time localisation dynamics in $\mathcal{H}^\pm$, as covered, for example, by the ``correlated volumes" approach mentioned at the end of Section~V-B. However, the exponentially-small correlations induced by the doublet pairs are specifically covered by $P_B(s)$. We thus expect the time scale associated to $K_W(\tau)= K_+(\tau)$ to be the Heisenberg time $t_H$. Saliently, $K_B(\tau)$ contains two different time scales. One rules the early-time dynamics and is also $t_H$, while the other rules the exponentially-large time doublet dynamics and is given by $t_D$.

\subsection{Nearest-level spacing distributions}

\begin{figure}
    \centering
    \includegraphics[width=0.99\linewidth]{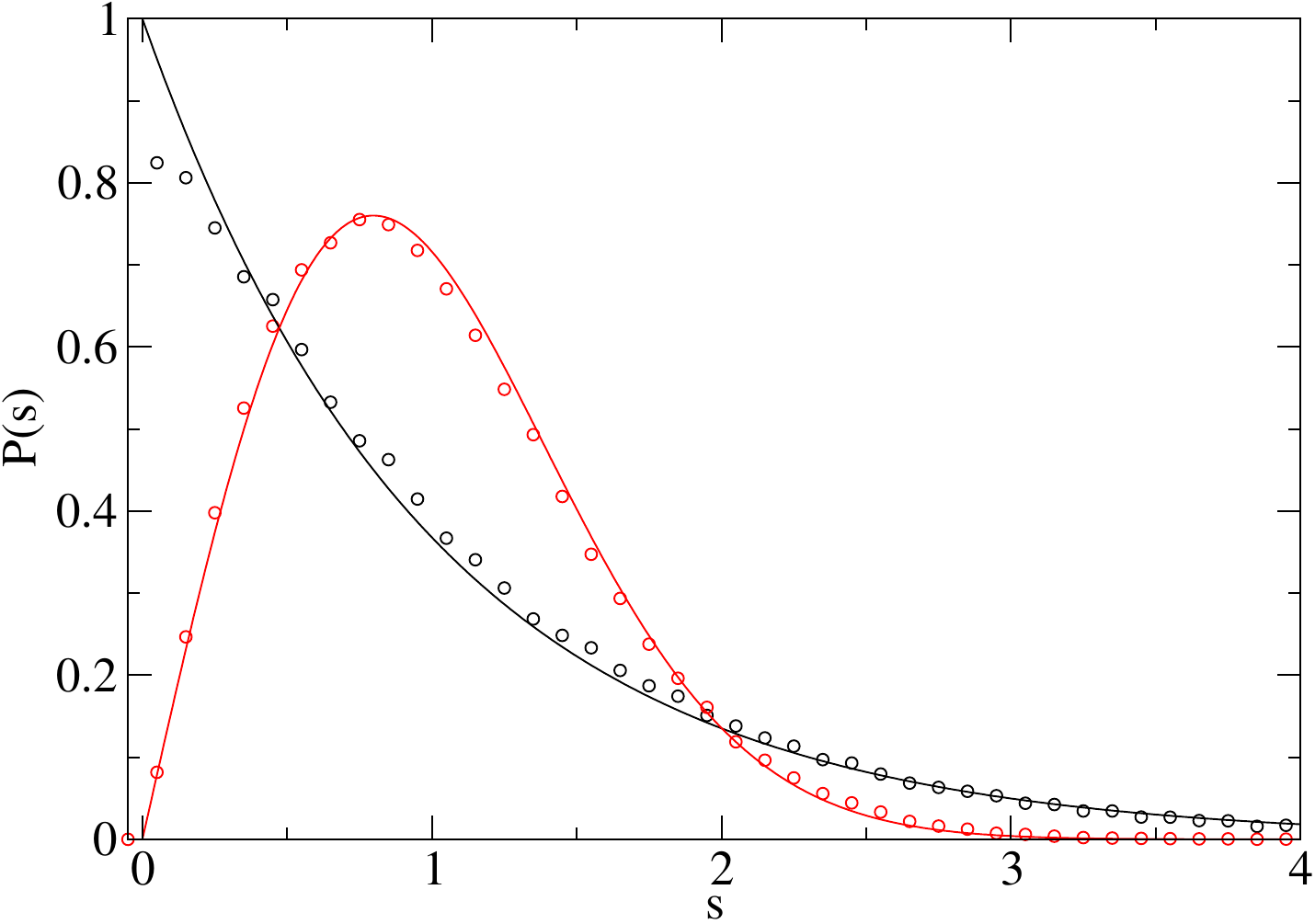}
    \caption{Circles: Nearest-level spacing distribution $P(s)$ obtained for the HRQKR model \eqref{eq:qkr_halk_kick_evolution_operator} with symmetric random phases in the even subspace $\mathcal{H}^+$ and with $\hbar_e=1$. The system size is $N=513$ and the data are averaged over $128$ disorder realizations. Since $\mathcal{H}^+$ contains $N/2$ states, quasi-energies have been rescaled by $2\Delta$ so that $\langle s\rangle=1$. The black circles have been obtained in the localised regime ($K=3$, $N_L=114$) while the red circles have been obtained in the metallic regime ($K=50$, $N_L=0.4$). Black solid line: Poisson distribution $P(s)=e^{-s}$. Red solid line: Wigner-Dyson distribution $P(s)=\frac12 \pi s\, e^{-\pi  s^2/4}$.}
    \label{fig:PdeSDemiSpectre}
\end{figure}

\begin{figure}
    \centering
    \includegraphics[width=0.99\linewidth]{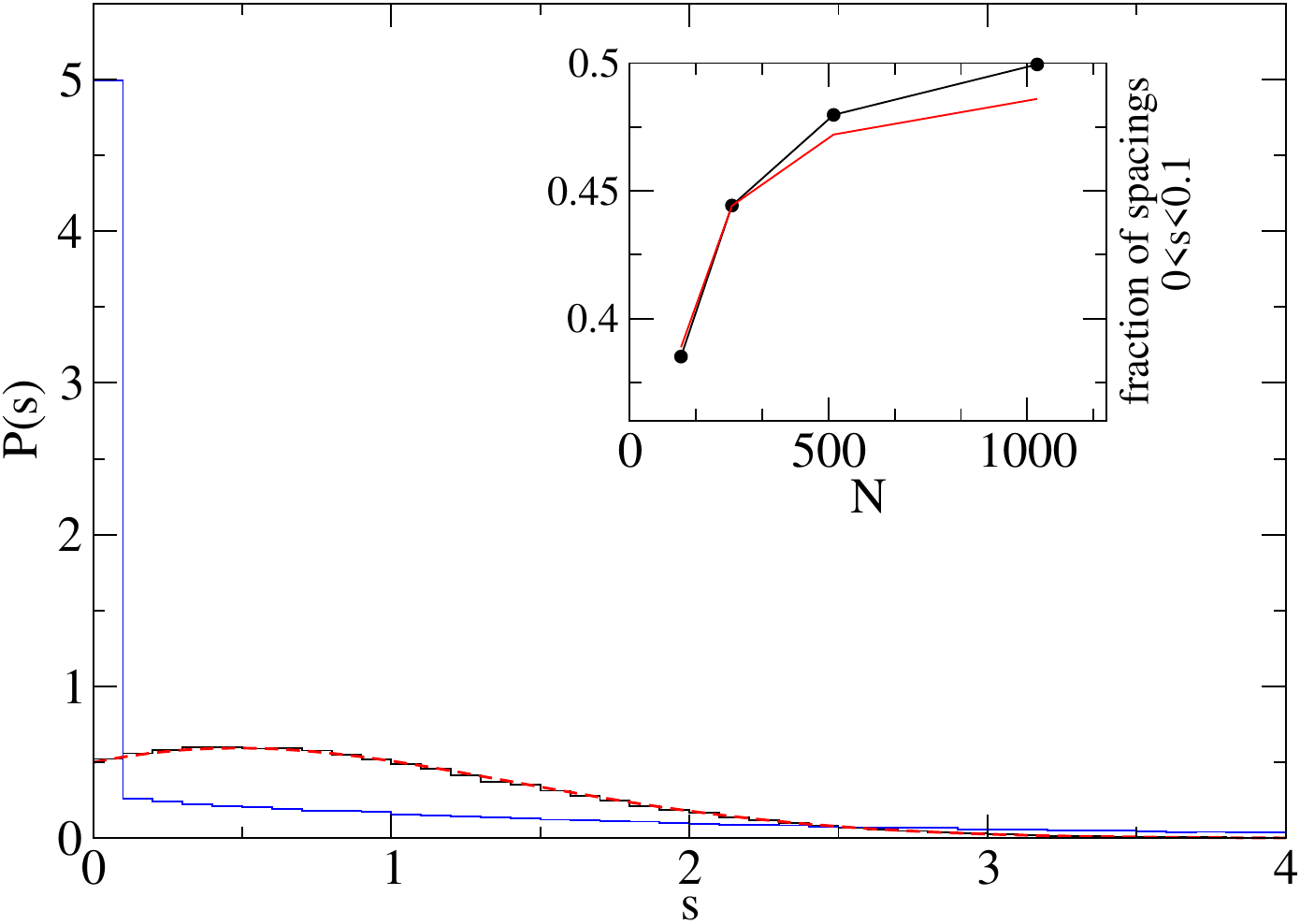}
    \caption{Main panel: Nearest-level spacing distribution $P(s)$ for the RQKR model \eqref{eq:qkr_halk_kick_evolution_operator} with symmetric random phases in the full Hilbert space $\mathcal{H}=\mathcal{H}^+\oplus\mathcal{H}^-$ and with $\hbar_e=1$. Black data (averaged over $128$ disorder realizations): Metallic regime $K=50, N=513$ and $N_L = 0.4$. Blue data (averaged over $256$ disorder realizations): Localised regime $K=3, N=1025$ and $N_L=228$. Quasi-energies are rescaled by $\Delta=2\pi/N$ so that $\langle s\rangle=1$. Dashed red line: theoretical prediction $P_{\text{mix}}(s)=\frac{\pi s}{8}e^{-\frac{\pi  s^2}{16}} \text{Erfc}(\frac{\sqrt{\pi}\,s}{4})+\frac12 e^{-\frac{\pi  s^2}{8}}$ obtained for the mixture of two 2 independent GOE spectra \cite{Giraud2022}. 
   Inset: Numerically-extracted fraction of spacings in the Shnirelman peak for $0\leq s\leq 0.1$ and in the localised case. Parameters are $K=3$ and $N=127$, $257$, $513$ and $1025$. For $N=1025$, the estimated fraction is $\sigma_S=49.95\%$, compatible with the fraction $49.78\%$ computed with Eq.~\eqref{eq:FracS} and $\xi_K/N = 2.3 \, 10^{-3}$ ($N_L=228$). Red curve: Fit of the data with the theoretical prediction $\sigma_S= 1/2-\xi/N$, Eq.~\eqref{eq:FracS}. A reasonable agreement is obtained for $\xi = 14.34$.
   }
    \label{fig:p(s)}
\end{figure}

\begin{figure}
    \centering
    \includegraphics[width=0.99\linewidth]{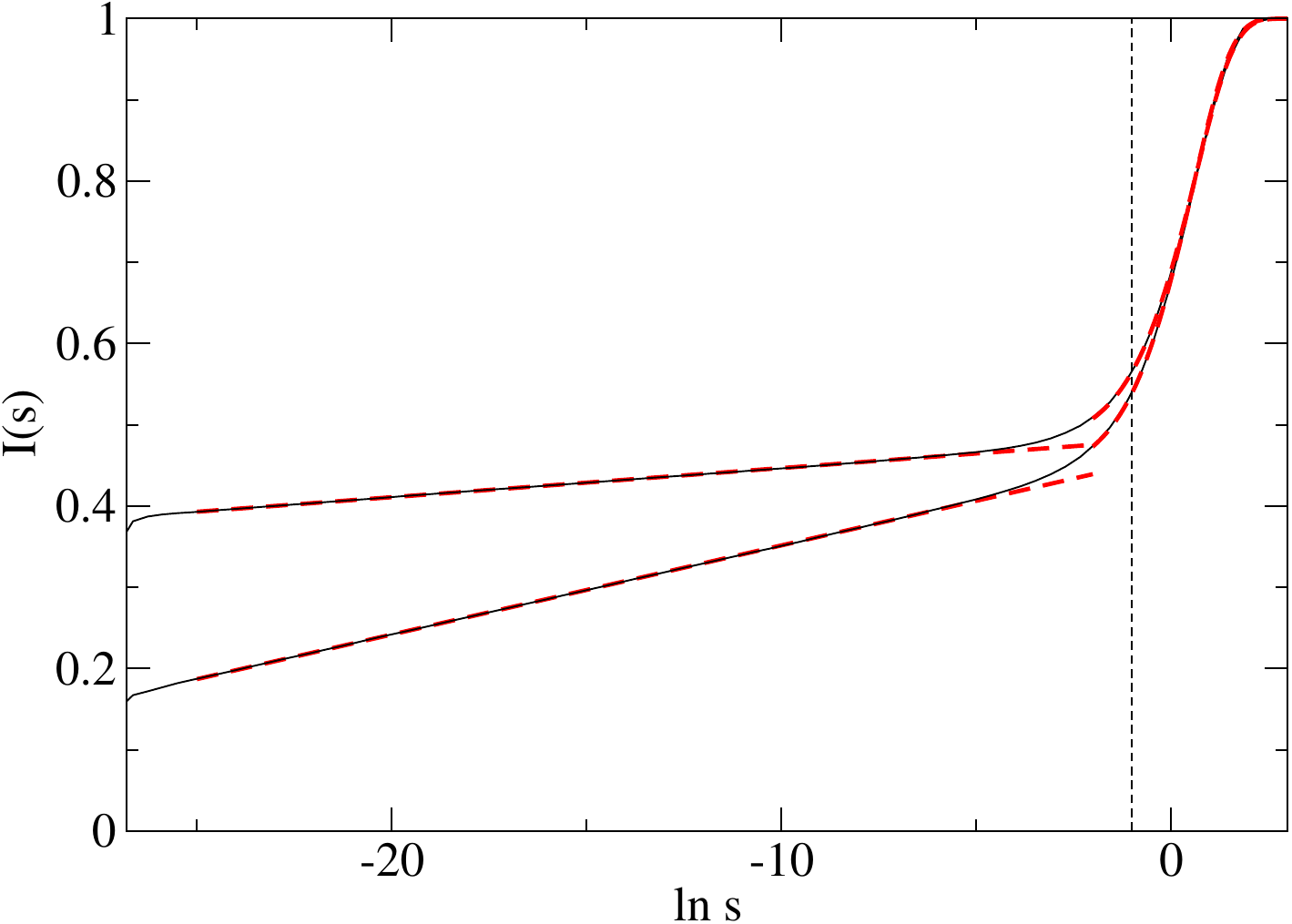}\\
     \includegraphics[width=0.99\linewidth]{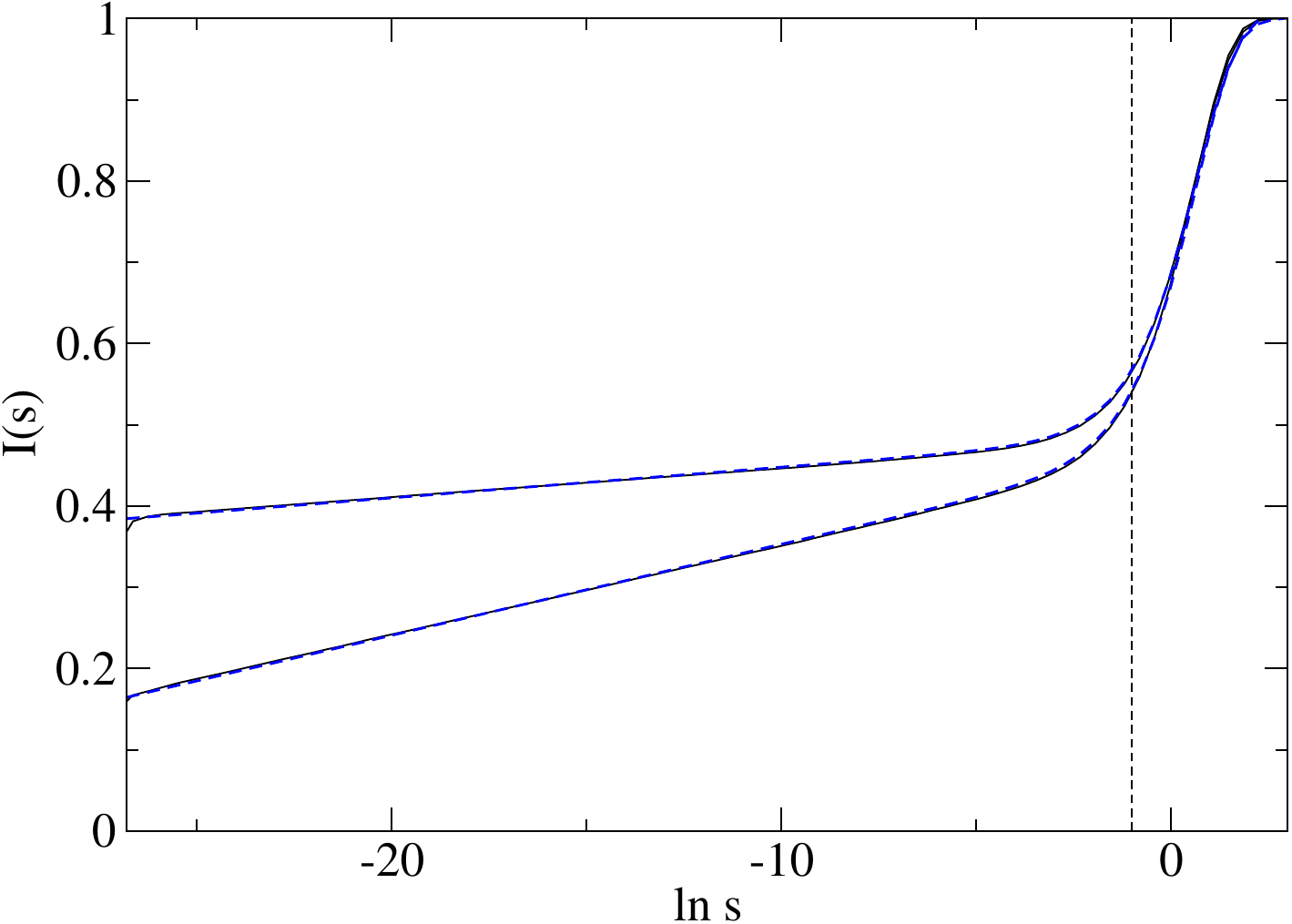}
   \caption{Integrated nearest-level spacing distribution $I(s)=\int_{0}^s P(s')ds'$ obtained for $\hbar_e=1$ as a function of $\ln s$ for $K=3$ (upper solid black curve) and $K=6$ (lower solid black curve). The system size is $N=1025$ and the data have been averaged over $256$ disorder realizations. Top panel: Separate fits for the linear and bell-shaped parts of the numerical data. The linear part for $K=3$ has been fitted over the interval $\ln s\in [-25,-5]$ with a straight line having a slope $\mu = 0.00358\simeq 3.67/N$ (upper red dashed line). The bell-shaped region around $\ln s =0$ has been fitted over the interval $\ln s \geq -2$ by the integrated Poisson distribution $I_P(s)=\int_0^s \sigma^2 e^{-\sigma s'} ds' = \sigma(1-e^{-\sigma s})$ plotted versus $\ln s$ (upper red dashed curve) where $\sigma$ is taken as a fit parameter.
%$I(s)= 1-\sigma e^{-\sigma \exp x}$. 
    We have found $\sigma \approx 0.53$, giving $\sigma_S=0.47$ (see main text). 
%The lower black curve has been obtained for $K=6$.
    The lower red dashed linear and curve fits for $K=6$ give $\mu=0.01097\simeq 11.24/N$, $\sigma\approx0.57$ and $\sigma_S = 0.43$. 
     These $\sigma_S$ values should be compared with the fractions of spacings in the interval $[0,0.1]$, which are almost $0.5$ and $0.46$, respectively. This systematic but small deviation is probably due to the arbitrariness in the choice of the spacing $s$ delineating the Shnirelman peak and the Poisson tail. The vertical dashed line marks $s_{\max} = e^{-1}$. Bottom panel: Fit of the data obtained for $K=3$ and $K=6$ using Eq.~\eqref{eq:IntP} with the constant term $1/2$ replaced by $a$ and ($\sigma,a$) as free parameters. We find $\sigma\approx0.504$, $a\approx0.4855$ for $K=3$ and $\sigma \approx 0.511$, $a = 0.465$ for $K=6$. As one can see the agreement with Eq.~\eqref{eq:IntP} is quite good. 
     }
    \label{fig:IdeSToutSpectreKickedRotorK3-6N1025}
\end{figure}

\begin{figure}
    \centering
    \includegraphics[width=0.99\linewidth]{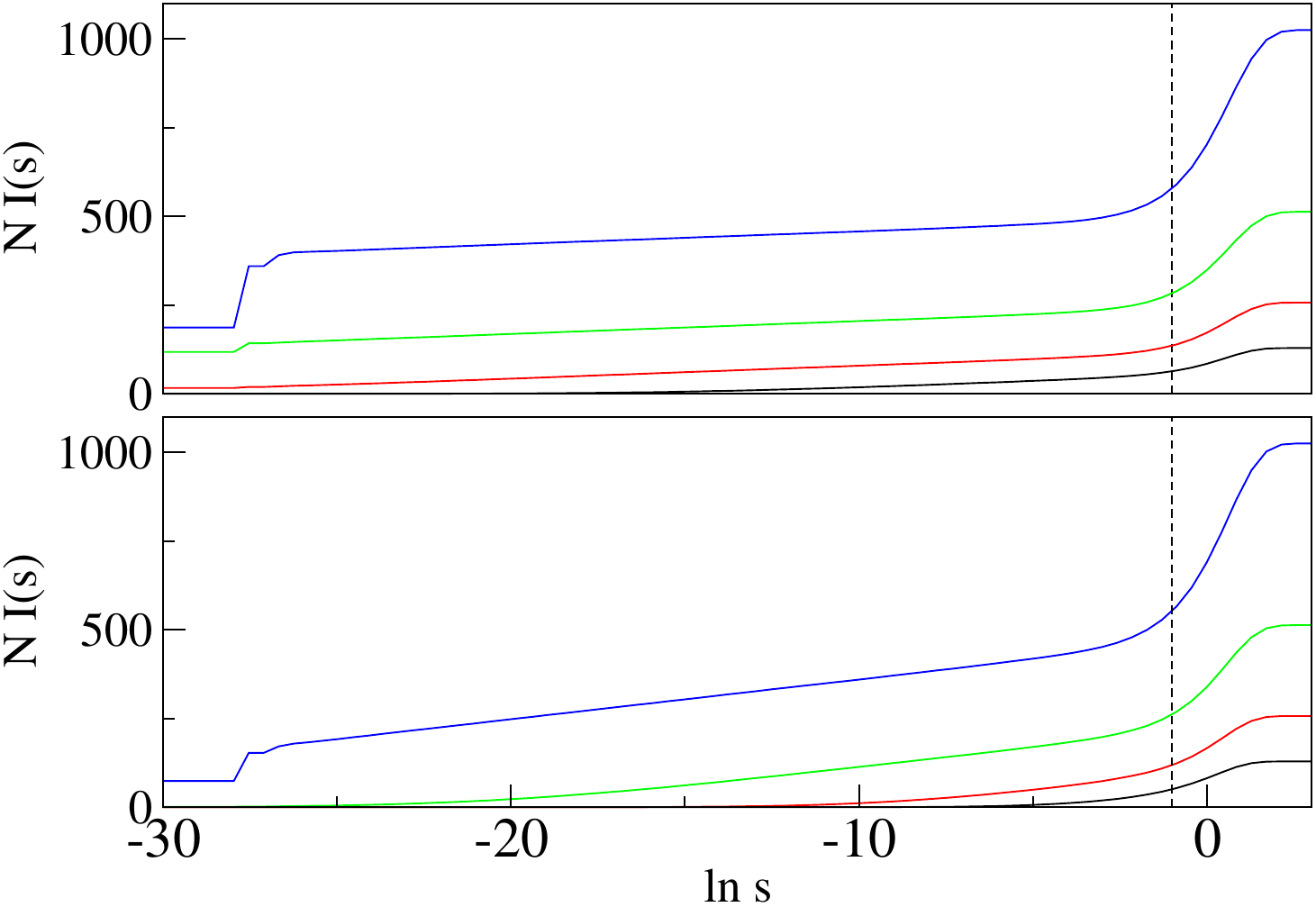}
    \caption{Rescaled integrated nearest-level spacing distribution $N I(s)=N\int_0^s P(s')ds'$ as a function of $\ln s$ at $\hbar_e=1$. Top panel: $K=3$. Bottom panel: $K=6$. In both panels, the curves are computed for system sizes   $N=129$ (black), 257 (red), 513 (green) and 1025 (blue). The linear region in $\ln s$ manifests only for $\xi/N$ small enough (localised regime). Our numerics show that the linear slopes of $NI(s)$ are indeed independent of $N$ since they are all the same in each panel. The slope of $NI(s)$ is given by $\xi$ solely. The vertical dashed line marks $s_{\max} = e^{-1}$.}
    \label{fig:IdeSToutSpectreKickedRotorK3-6N7-10}
\end{figure}

In Fig.~\ref{fig:PdeSDemiSpectre}, we plot the nearest-level spacing distribution $P(s)$ obtained in $\mathcal{H}^+$ (the results are identical for $\mathcal{H}^-$). Its normalization is set to $\int_0^{\infty} P(s) ds = 1$ and $\langle s \rangle = \int_0^{\infty} s P(s) ds =1$. The data are well reproduced by the expected Poisson distribution in the localised regime ($N_L \gg 1$) and by the expected Wigner-Dyson distribution derived from  the Gaussian Orthogonal Ensemble (GOE) of Random Matrix Theory (RMT) in the metallic regime ($N_L \lesssim 1$). From this, we can infer that the time scale relevant to $C_W(x_0,t)$ is the Heisenberg time of the system, either $t_H=\xi$ in the localised regime, or $t_H = N$ in the metallic regime. 

In Fig.~\ref{fig:p(s)}, we now plot the nearest-level spacing distribution $P(s)$ obtained in the localised ($N_L\gg 1$) and metallic ($N_L \lesssim 1$) regimes but for the full Hilbert space $\mathcal{H}=\mathcal{H}^+\oplus\mathcal{H}^-$. In the metallic regime, the data are well reproduced by the mixture of two independent GOE spectra (one is $\mathcal{S}^+$, the other $\mathcal{S}^-$) having the same statistical properties \cite{Giraud2022}.
In the localised regime however, we find a Poisson-like behaviour at sufficiently large spacings and a peak developing at $s\to 0$. This peak, reminiscent of the Shnirelman peak discussed in \cite{chirikov1995shnirelman}, reveals that a macroscopic number of pairs of levels, one from $\mathcal{S}^+$, the other from $\mathcal{S}^-$, are accumulating at $s$ very small. These are the doublet states originating from the parity-induced hybridization mechanism mentioned earlier. Their effect on the very large time behaviour of $C_B(x,t)$ is of paramount importance.

The Shnirelman peak contribution $P_S(s)$ to $P(s)$ is described by the $s\ll 1$ limit of $P_B(s)$. The mean energy of the doublet states being uniformly distributed over $2\pi$, we expect
\begin{equation}
    P_{S}(s) \sim \frac{1}{N}\sum_{k=\xi}^{M} \delta(s - e^{-k/\xi}) \sim \frac{1}{2N_Ls} \ \mathbbm{1}_{[s_{\min},s_{\max}]},
    \label{eq:1/s}
\end{equation}
where the last term has been obtained by approximating the sum by an integral. Here $s_{\max}=1/e$ and $s_{\min} = e^{-N_L}$. This $1/s$ scaling is consistent with the one found in \cite{chirikov1995shnirelman}.
As we have seen, the Shnirelman peak is due to hybridized eigenstates having symmetric but distant enough localisation centres, excluding {\it de facto} the central localisation box. We thus expect the number $N_S$ of states participating to the Shnirelman peak to scale like the number of (positive) localisation centres outside the central box. We thus get the estimate $N_S \sim M-\xi = (N_L-1)\,\xi$. The corresponding fraction of states is thus
\begin{equation}
    \sigma_S=\frac{N_S}{N} = \frac{(N_L-1)}{2N_L} = \frac{1}{2}-\frac{\xi}{N},
    \label{eq:FracS}
\end{equation}
which is exactly the total weight of $P_S(s)$ given by its integral.
As noted in \cite{chirikov1995shnirelman}, $P(s)$ also features a Poisson-like contribution $P_P(s) = \sigma^2\exp(-\sigma s)$ where 
\begin{equation}
    \sigma=1-\sigma_S = \frac{N_L+1}{2N_L} = \frac{1}{2}+\frac{\xi}{N}. \label{eq:sigma}
\end{equation}
The weight of $P_P(s)$, given by its integral, is precisely $\sigma$ such that the total weight of $P(s)$ is indeed $\sigma+\sigma_S=1$. In the metallic regime $N_L \lesssim 1$, the Shnirelman peak disappears since $\sigma_S \to 0$ and $\sigma \to 1$. In this case, though still present, the parity-induced hybridization mechanism cannot lead to exponentially close quasi-energies and $P(s)$ originates from the mixing of two independent, but statistically identical, random matrix ensembles associated to each parity sector, as seen in Figs.~\ref{fig:PdeSDemiSpectre} and \ref{fig:p(s)} .  In the localised regime $N_L \gg 1$, $\sigma_S = \sigma = 50\%$ and the two contributions to $P(s)$ have equal weights.

To check these predictions, we follow \cite{chirikov1995shnirelman} and numerically compute the integrated nearest-level spacing distribution $I(s)= \int_0^s ds' P(s')$ and confront our data to the theoretical prediction inferred from the previous considerations, namely
\begin{equation}
    I(s)= \frac{1}{2}+ \sigma (1-e^{-\sigma s}) + \frac{2\sigma-1}{2} \ln s.
    \label{eq:IntP}
\end{equation}
At $s\ll 1$, the behaviour of $I(s)$ is completely dominated by the $\ln s$ term. To highlight this fact, we plot $I(s)$ as a function of $\ln s$ for two values of $K$ in the localised regime, see Fig.~\ref{fig:IdeSToutSpectreKickedRotorK3-6N1025}, and we use Eq.~\eqref{eq:IntP} as a fitting curve to extract $\sigma$. The predicted slope of the $\ln s$-linear region in Eq.~\eqref{eq:IntP} is $(2N_L)^{-1}=\xi/N$, see Eq.~\eqref{eq:sigma}. In Fig.~\ref{fig:IdeSToutSpectreKickedRotorK3-6N7-10}, we plot $NI(s)$ against $\ln s$ for various system sizes at fixed $K$. We observe that the slopes of the linear regions are all the same, confirming the $N^{-1}$ scaling in Eq.~\eqref{eq:1/s}.

\section{Dynamical behaviours of the CFS and CBS peaks in the different regimes} 

\subsection{Dynamics at the Heisenberg time $t_H$ scale}

\begin{figure*}
    \centering
    \includegraphics[scale=0.5]{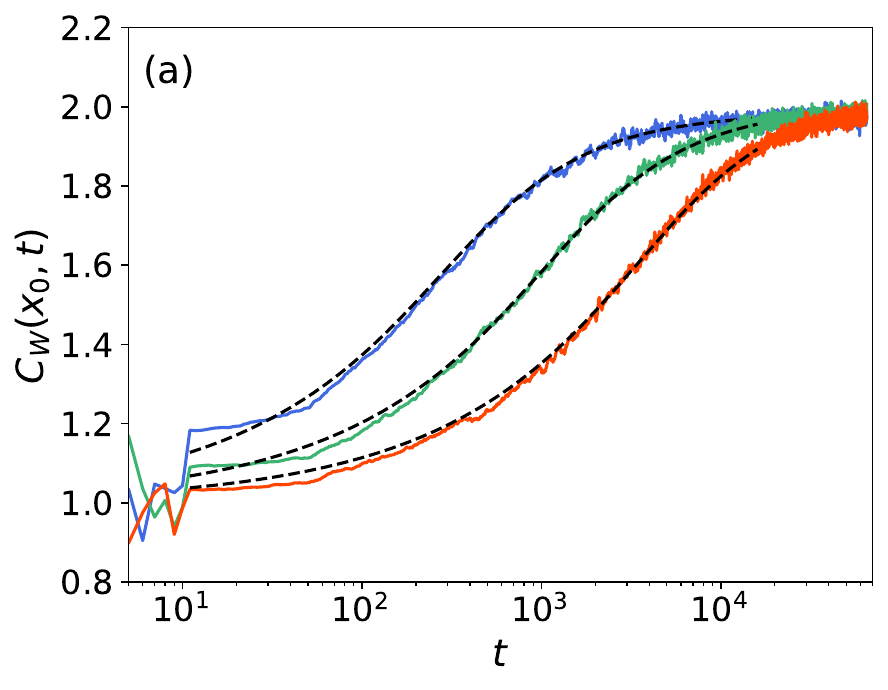}
    \includegraphics[scale=0.5]{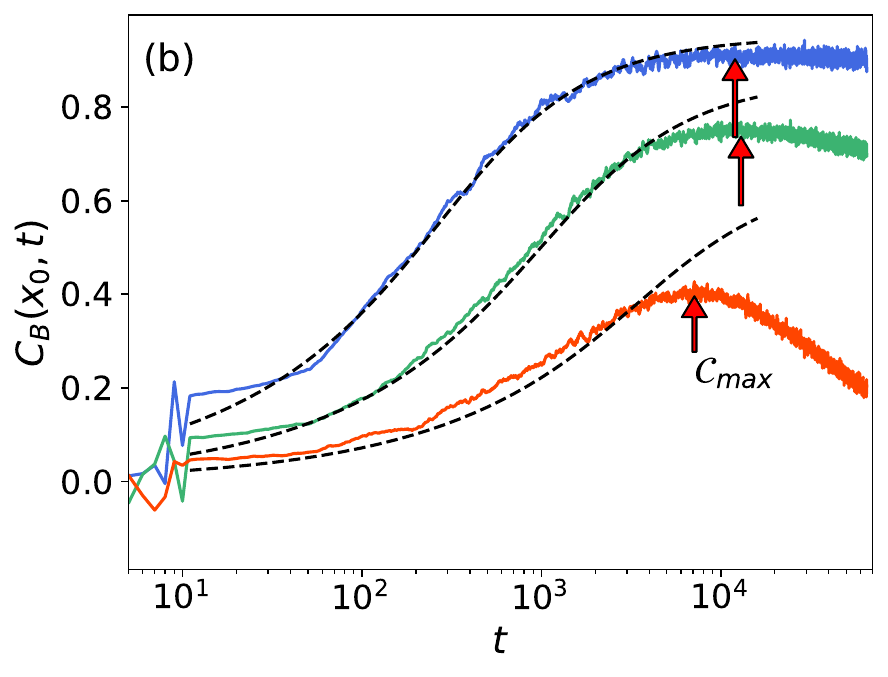}\\
    \includegraphics[scale=0.245]{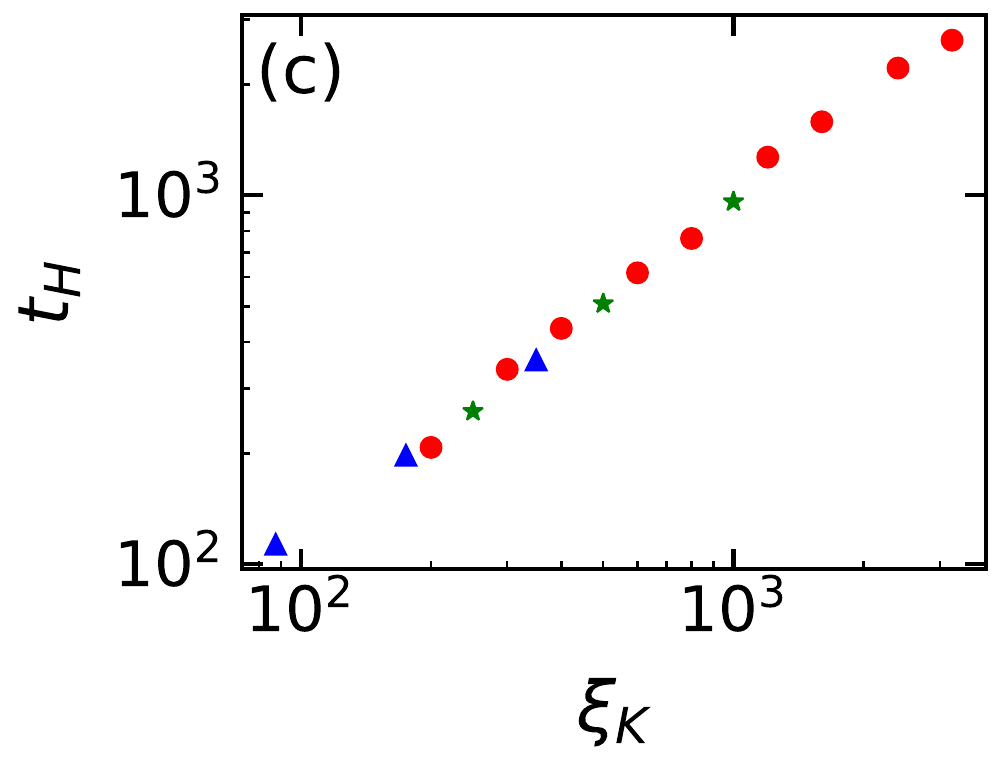}
    \includegraphics[scale=0.245]{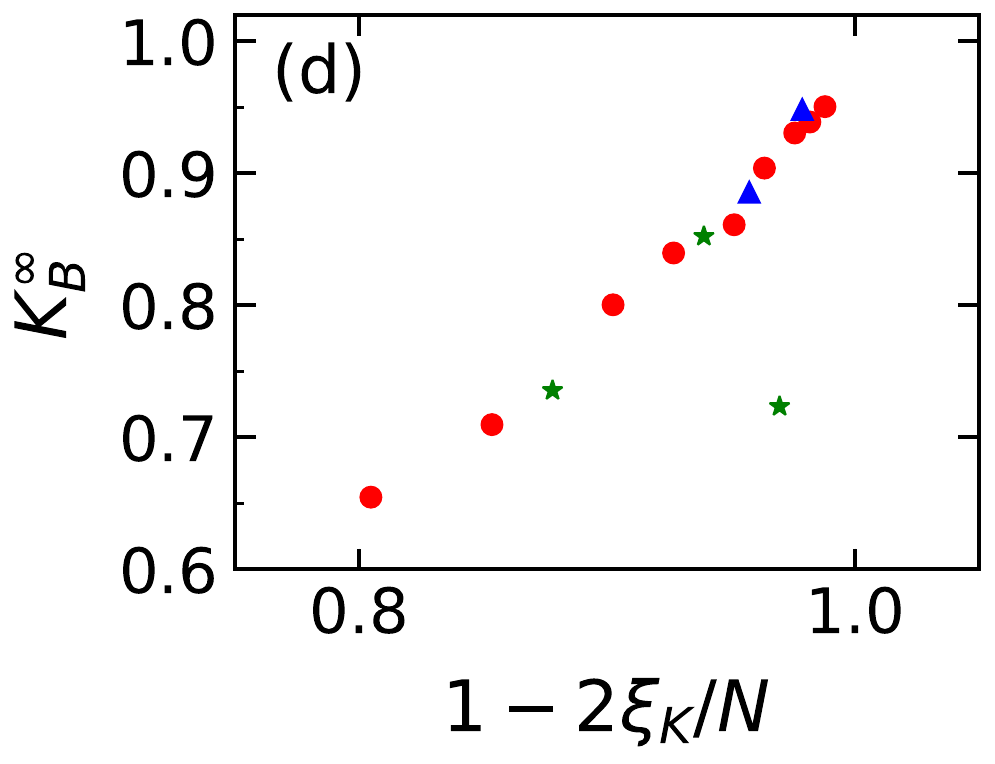}
    \includegraphics[scale=0.245]{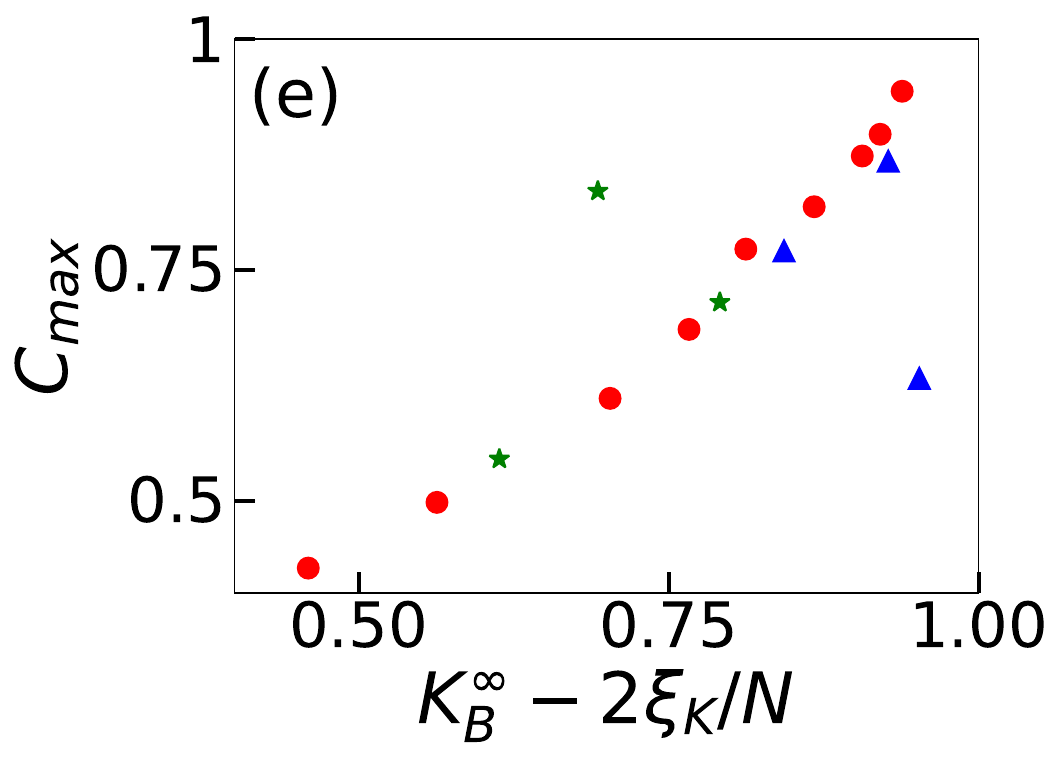}
    \includegraphics[scale=0.245]{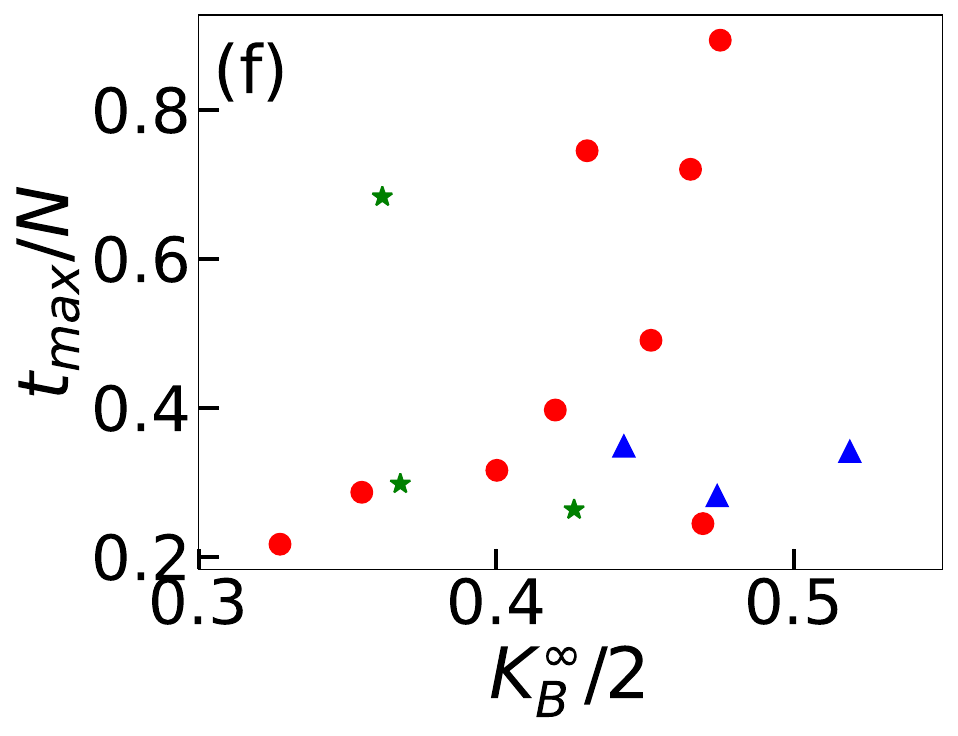}
    \caption{Panel (a): Temporal behaviour of $C_W(x_0,t)$. Panel (b): Temporal behaviour of $C_B(x_0,t)$. For both panels, the dynamics is computed up to time $t_{\mathrm{sim}} = 65\hspace{2pt}000$ with $\hbar_e = 1$ and $M=16\hspace{2pt}384$ (lattice size $N = 2M+1= 32\hspace{2pt}769$). The localisation length $\xi_K\equiv (K/2\hbar_e)^2$ is the varying parameter, ranging from $87.5$ to $3200$. To maintain clarity, we only show the curves obtained for $K = 28$ (blue curves, $\xi_K = 196$), $K = 56$ (green curves, $\xi_K = 784$) and $K = 112$ (red curves, $\xi_K = 3136$). The dashed black lines are the analytical predictions $C_W(x_0,t) = 1+K^\infty_W \, M(t/t_H)$ and $C_B(x_0,t) = K^\infty_B \, M(t/t_H)$ obtained from $t = 10$ to $t = t_{\mathrm{sim}}/4$ with the function $M(\tau)$  given by Eq.~\eqref{eq:MickFunc} and $K^\infty_{W}$, $K^\infty_{B}$ and $t_H$ as fit parameters (see main text). 
    Panel (c): Localisation time $t_H$ plotted against $\xi$ showing, as expected, that $t_H \sim \xi$ in the localised regime. Since the early-time dynamics does not involve doublets, the localisation time $t_H$ is the same for both $C_W(x_0,t)$ and $C_B(x_0,t)$, as seen in the plots. 
    Panel (d): fit parameter $K^\infty_B$, which varies between $0.6$ and $1$ as localisation gets stronger. The fit parameter $K^\infty_W \sim 1$ for all values (data not shown). 
    Panel (e): Plot of the maximum value $C_{\max}$ reached by $C_B(x_0,t)$, and materialized by the red arrows in panel (b), as a function of $K_B^\infty-2\xi_K/N$ showing, as expected, that $C_{\max} \approx K_B^\infty-2\xi_K/N$. Since $C_{\max}$ can exhibit sizeable fluctuations, the values shown are obtained by averaging the contrast time plots over 40 time points. Panel (f): The characteristic time $t_{\max}$ where $C_B(x_0,t)$ reaches its maximum value $C_{\max}$ is plotted as a function of $K_B^\infty/2$. The data loosely agree with the expected behaviour $t_{\max}/N \sim K_B^\infty/2$. Do note however that extracting $t_{\max}$ from the data becomes difficult for $N_L \gg 1$ and $K^\infty_B \to 1$ (localised regime). Panels (c-f): The red circle symbols have been obtained for $K = 28, 35, 40, 49, 56, 69, 80, 98, 112$ and $M=16\hspace{2pt}384$, the green star symbols for $K = 31.62,44.72,63.25$ and $M = 8\hspace{2pt}192$, and the blue triangle symbols  for $K = 18.71,26.46,37.42$ and $M = 4092$. 
    }
\label{fig:n+time}
\end{figure*}

In Fig.~\ref{fig:n+time}(a,b), we plot $C_W(x_0,t)=1+K_{+}(t)$ and $C_B(x_0,t) \sim K_B(t)$ as a function of time for several kick strengths $K$ at fixed system size $N$. We expect the early-time dynamics of the system to be immune to the parity-induced doublet structure of the spectrum and thus to be controlled by the usual level repulsion aspects. As such, we anticipate that the early-enough temporal dynamics of both $K_{+}(t)$ and $K_B(t)$ is governed by the {\it same} function 
\begin{equation}
M(\tau)=[I_0(2/\tau)+I_1(2/\tau)]e^{-2/\tau},
\label{eq:MickFunc}
\end{equation}
where $I_0$ and $I_1$ are the modified Bessel functions of order 0 and 1 \cite{PhysRevB.97.041406}. As can be seen in Fig.~\ref{fig:n+time}(a,b), the curves $C_W(x_0,t) = 1+ K^{\infty}_{W} \, M(t/t_H)$ and $C_B(x_0,t) = K^\infty_{B} \, M(t/t_H)$ fit well with the numerical data, which allows us to extract $K^\infty_{W}$,  $K^\infty_{B}$ and $t_H$ as a function of the parameters. We find that, as expected in the localisation regime, $K^\infty_{W} \sim 1$ and $t_H=\xi$ (see Fig.~\ref{fig:n+time}c) over the whole range of parameters explored. $K^\infty_{B}$ varies between $0.6$ and $1$ as localisation gets stronger, see Fig.~\ref{fig:n+time}d.

\subsection{Long-time dynamics}

\begin{figure}
    \centering
    \includegraphics[scale = 0.55]{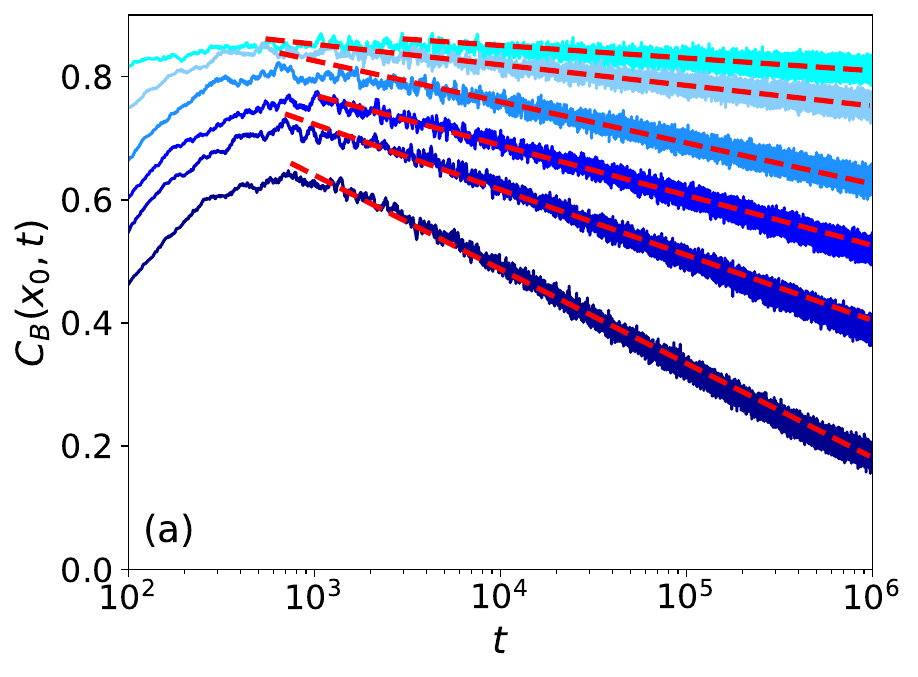}
    \includegraphics[scale = 0.23]{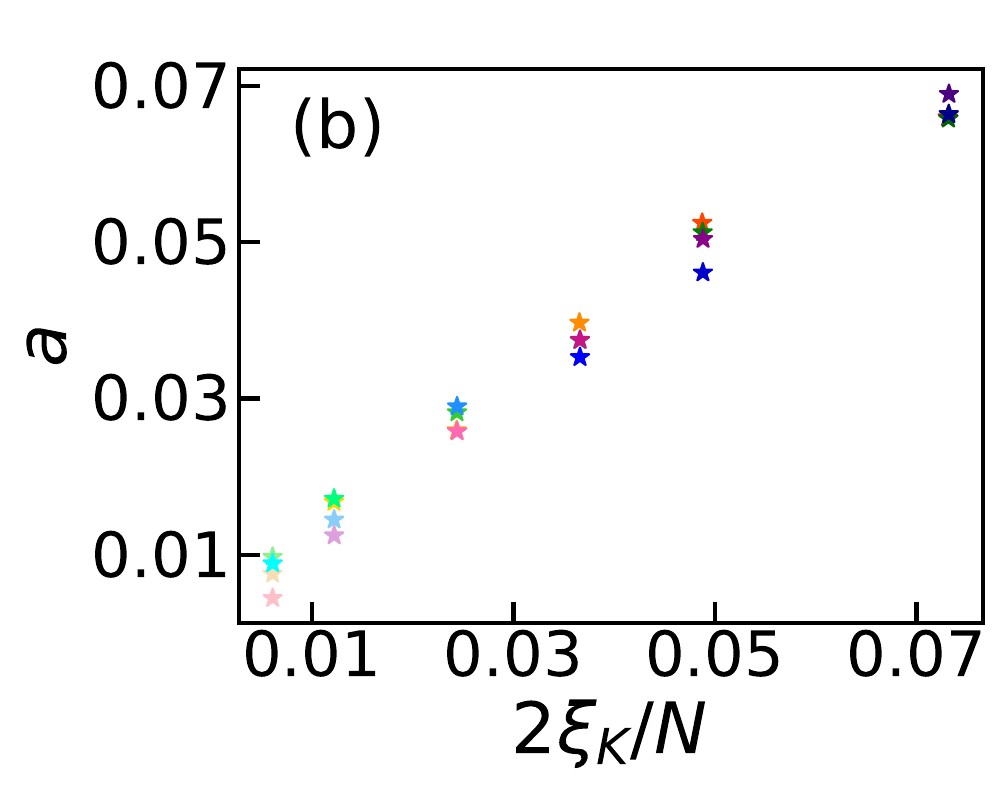}
    \includegraphics[scale = 0.23]{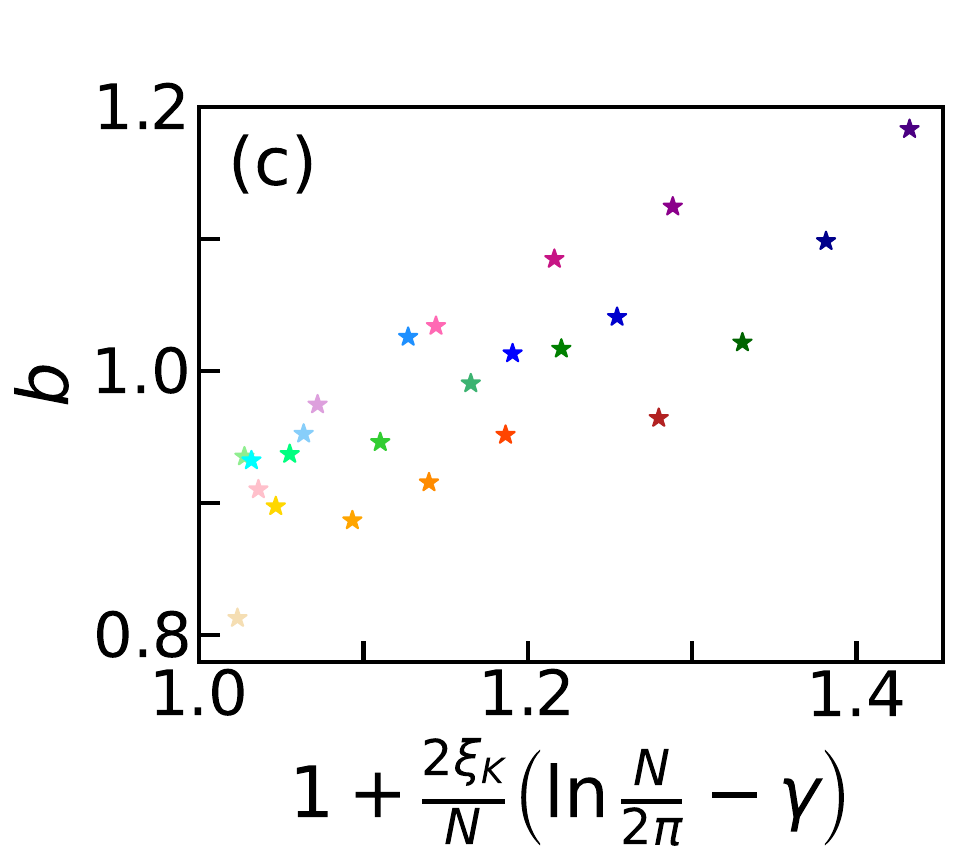}
    \caption{Panel a: Logarithmic relaxation of $C_B(x_0,t)$ after the time $t_{\max}$.
    %when $C_{\max} \equiv C_B(x_0,t_{max}) = \max[C_B(x_0,t)]$. 
    The blue curves represent a fixed $M=1024$ (system size $N =2M+1= 2049$) with localisation lengths $\xi_K$ ranging from $6.25$ to $75$ (lighter to darker blue). The red dotted lines are logarithmic fits for times after $t_{\max}$, of the form $C_B(x_0,t)=-a \ln t + b$, see Eq.\eqref{eq:ab} (the fit parameter $b$ just amounts to a mere vertical translation of the curves). Panel b: Slope $a$ of the logarithmic fit as a function of $2\xi_K/N$. Our numerical data confirm the theoretical scaling $a \sim 2\xi_K/N$ predicted by Eq.~\eqref{eq:ab}. Panel c: constant $b$ of the logarithmic fit, showing that it is of order 1 in the variable $1 + \frac{2\xi}{N} \left(\ln \frac{N}{2 \pi} - \gamma\right) $.
    As in the other figures, these data are computed with $\hbar_e = 1$ and are averaged over 1080 disorder realizations. Since they exhibit significant fluctuations, the values shown are obtained by averaging the contrasts over $40$ consecutive time points.}
\label{fig:Logbehaviour}
\end{figure} 

\begin{figure}
    \centering
    \includegraphics[scale = 0.55]{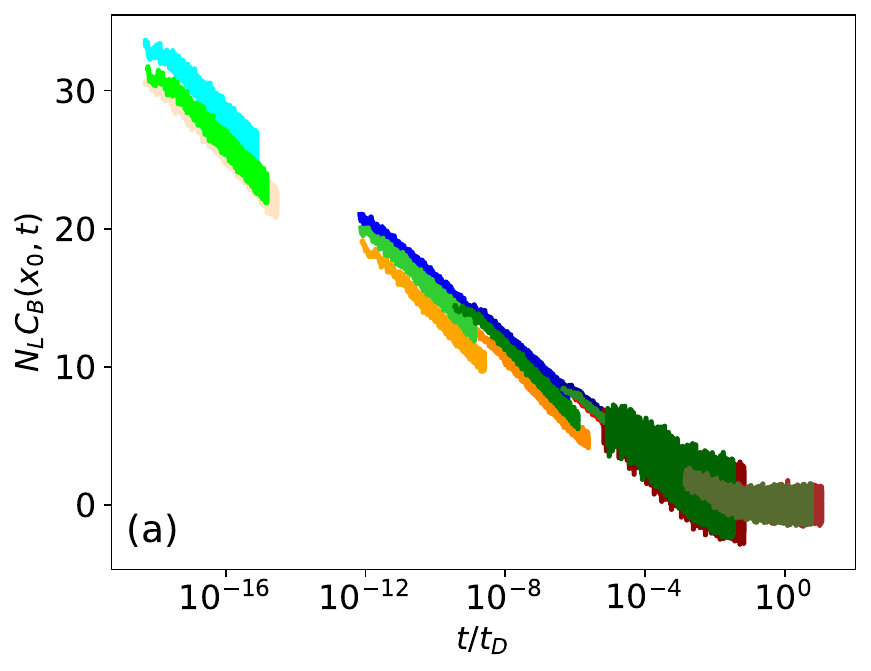}
    \includegraphics[scale = 0.55]{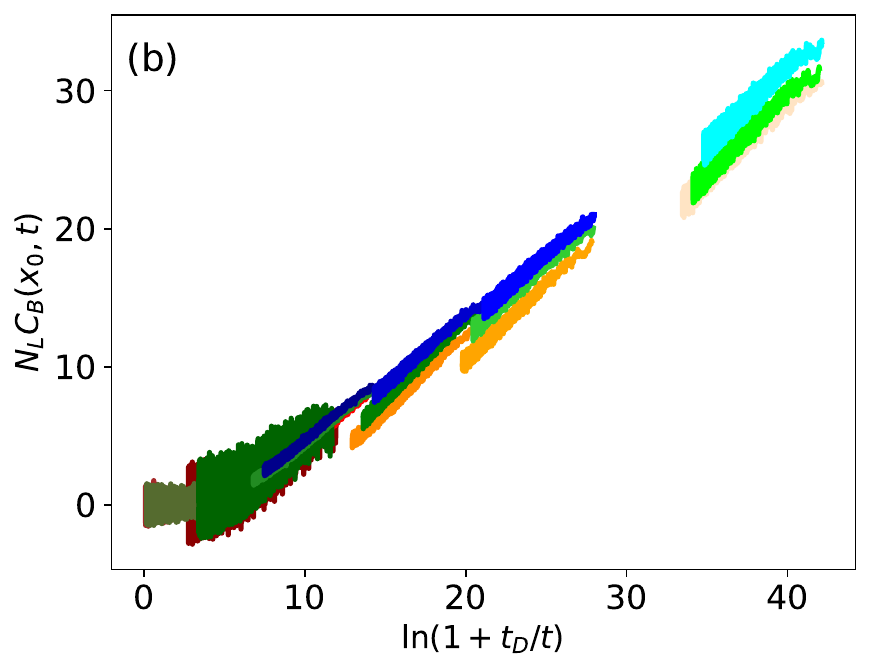}
    \caption{ Scaling behaviour of the CFS contrast $N_L [C(x_0,t)-2] \approx N_L C_B(x_0,t)$, at large times $t\gg t_{\max}$, versus $t/t_D$, showing a behaviour identical to logarithmic relaxation and aging in glassy systems, see main text and Eq.~\eqref{eq:logrelax}. In this analogy, $t_D = N \exp(N_L) = N \exp[N/(2 \xi_K)]$, with $\xi_K = (K/2\hbar_e)^2$, plays the role of the so-called waiting time. 
    The blue curves represent a fixed $M=1024$ (system size $N =2M+1= 2049$) with localisation lengths $\xi_K= 25, 37.5, 50, 75$ (lighter to darker blue). The green curves represent a fixed $M=512$ (system size $N =2M+1= 1025$) with localisation lengths from $\xi_K= 12.5$ up to $100$ (lighter to darker green). The red curves represent a fixed $M=256$ (system size $N =2M+1= 513$) with localisation lengths from $\xi_K= 6.25$ up to $50$ (lighter to darker red). As in the other figures, these data are computed with $\hbar_e = 1$ and are averaged over 1080 disorder realizations. Since they still exhibit significant fluctuations, the values shown are obtained by averaging further over $40$ consecutive time points.}
\label{fig:ScaCBtD}
\end{figure} 

As seen in Fig.~\ref{fig:n+time}a, $C_W(x_0,t) \to C_\infty \approx 2$ after a few Heisenberg times of the system ($t_H \sim \xi$ in the localised regime and $t_H \sim N$ in the metallic regime). This means that the intra-sector dynamics ``quickly" saturates and reaches its stationary intra-sector limit. The remaining CFS/CBS dynamics $C(\pm x_0,t)$ at times $t \gg t_H$ is thus encapsulated by $C_B(x_0,t)$ and thus dictated by the Shnirelman peak $P_S(s)$, Eq.~\eqref{eq:1/s}.

The contribution $C_B(x_0,\tau)$ is related to the spectral form factor via Eq.~\eqref{eq:KDminus}. As discussed below Eq.~\eqref{twopointfunction}, its $\tau\to\infty$ behaviour is governed by the $s\to 0$ limit of its Fourier transform, which itself is dominated at $s\to 0$ by the nearest-neighbour spacing. Taking the Fourier transform of Eq.~\eqref{eq:1/s} gives
\begin{eqnarray}
    C_B(x_0,\tau) =K_B(\tau) \sim \frac{2\xi}{N} \, \int_{s_{\min}}^{s_{\max}} ds \, \frac{\cos(s\tau)}{s} \\= \frac{1}{N_L} \,[\textrm{Ci}(s_{\max}\tau)-\textrm{Ci}(s_{\min}\tau)],
    \label{eq:KD}
\end{eqnarray}
where $\mathrm{Ci}(x) = - \int_x^\infty ds\, \cos s /s$ is the cosine integral function and we recall that $s_{\min}=e^{-N_L}$ and $s_{\max}=1/e$, with $N_L$ given by Eq.~\eqref{eq:NB}.

For $s_{\max}\tau\ll 1$ (and consequently $s_{\min}\tau\ll 1$), we can use the asymptotics $\mathrm{Ci}(x\ll1) \sim \gamma +\ln x$ (with $\gamma\approx 0.577$ the Euler-Mascheroni constant) for both cosine integral functions in Eq.~\eqref{eq:KD} and find $C_B(x_0,\tau \ll 1) \approx 2\sigma_S$, with $\sigma_S$ given by \eqref{eq:FracS}. For $ s_{\min}\tau \ll 1 \ll s_{\max}\tau$, we can use the asymptotics 
$\mathrm{Ci}(x\gg 1) \sim \mathrm{sinc}(x)$ for the first cosine integral function, where  $\mathrm{sinc}(x)$ is the sine cardinal function, while keeping the asymptotics $\mathrm{Ci}(x\ll1)$ for the other one. Since $\mathrm{sinc}(s_{\max}\tau)$ quickly vanishes before hitting $s_{\min}\tau \lesssim 1$, we can safely write
\begin{equation}
    C_B(x_0,\tau) = \, K_B(\tau) \approx - \frac{(\gamma + \ln s_{\min} \tau)}{N_L}.
    \label{eq:DFormFactor}
\end{equation}
Using $s_{\min}=e^{-N_L}$ and $\tau=t\Delta = t 2\pi/N$, we can also write 
\begin{equation}
    C_B(x_0,t) \approx  1 + \frac{2\xi}{N} \left(\ln \frac{N}{2 \pi} - \gamma\right)- \frac{2\xi}{N} \,\ln t\,,
    \label{eq:ab}
\end{equation}
which shows that $C_B(x_0,t)$ is logarithmically decreasing while staying positive in this range of times. Figure \ref{fig:Logbehaviour} confirms these predictions. It displays in panel (a) the large-time $t\gg t_{\max}$ behaviour of $C_B(x_0,t)$ and logarithmic fits of the form $C_B(x_0,t) = -a \ln t + b $, where the fitting parameters are shown in panels (b) and (c). The data confirm that $a \approx 2 \xi/N = N_L^{-1}$ and $b \approx 1$ for the localised regime considered $N_L \gg 1$.

When $s_{\min}\tau \gtrsim 1$, the doublets are fully resolved and the logarithmic behaviour is superseded by 
\begin{equation}
C_B(x_0,\tau) = K_B(\tau) \sim \frac{\mathrm{sinc}(s_{\max}\tau)-\mathrm{sinc}(s_{\min}\tau)}{N_L},
\label{eq:DFormFactorlongtimes}
\end{equation}
where we have used both large-argument limiting behaviours of the cosine integral functions.
%\new{(GL: there must be a constant missing which would make $C_B=K_B \geq 0$ - we do claim that $C_B\geq0$.)} 
We recover, at exponentially-large times, the limiting behaviour
$K_B(\tau)\to 0$ that we had inferred from previous considerations.

The logarithmic relaxation of the CFS contrast Eq.~\eqref{eq:DFormFactor} is the central result of our paper. Its behaviour is strongly reminiscent of the slow dynamics found in glasses  \cite{bouchaud1992, bouchaud1998out, amir2012relaxations}. This logarithmic relaxation has been experimentally observed in different glassy systems such as the electron glass \cite{PhysRevLett.92.066801, pollak2013electron, PhysRevB.88.085106} or a crumpling paper \cite{PhysRevLett.88.076101}. After perturbing such a system for an age $t_w$, the relaxation that follows behaves as \cite{amir2012relaxations}
\begin{equation}
    S(t,t_w) \propto \ln (1+t_w/t) \; ,
    \label{eq:logrelax}
\end{equation}
where $S$ is the physical observable measured, for example the conductance in an electron glass experiment. This type of time behaviour can be also identified in the dynamics of the CFS contrast in our system by choosing $S \equiv N_L [C(x_0,t)-2] \approx N_L C_B(x_0,t) $ (since $C_W(x_0,t) \approx 2$ at large times $t\gg t_H$) and $t_w \equiv t_D = N \exp(N_L)= N \exp[N/(2 \xi)]$. Indeed, recalling that $\tau=t\Delta$ with $\Delta=2\pi/N$ and using Eqs.~\eqref{eq:DFormFactor} and \eqref{eq:DFormFactorlongtimes}, we have 
\begin{equation}
    N_L [C(x_0,t)-2] \approx  \begin{cases}
        \ln t_D/t\; \text{ for }t_{\max} \ll t\ll t_D \\
        t_D/t \; \text{ for } t\gg t_D \;,
    \end{cases} \label{eq:scaCBtD}
\end{equation}
where $C(x_0,t)$ is given in Eq.\eqref{eq:C}. This is precisely the behaviour of logarithmic relaxations characterizing several different types of glassy systems \cite{amir2012relaxations, Amir2010}. 

Equation \eqref{eq:scaCBtD} indicates a non-trivial scaling property of the CFS contrast in terms of $t/t_D$. This scaling behaviour is illustrated in Fig.~\ref{fig:ScaCBtD}.
The behaviour that we observe actually indicates a phenomenon of aging. The relaxation depends explicitly on the ``waiting time'' $t_D$: the longer the ``perturbation'', the slower the logarithmic decay.  

The mechanism underlying this striking analogy relies on the power-law distribution 
of the nearest-level spacings in the Shnirelman peak, $P_S(s) \sim 1/s$, 
see Eq.~\eqref{eq:1/s} and \cite{chirikov1995shnirelman}. This can be interpreted as yielding a broad distribution 
of relaxation rates $s$ for the doublets. 
A simple model for logarithmic relaxation in glassy systems
\cite{amir2012relaxations} indeed consists of a collection of many independent modes 
that relax with rates $\lambda$ distributed as $P(\lambda) \sim 1/\lambda$. 
The observable $S(t)$ at time $t$ after the perturbation can then be written as
\begin{eqnarray}
    S(t) &\approx& S(0) \int_{\lambda_{\min}}^{\lambda_{\max}} 
    e^{-\lambda t} P(\lambda)\, d\lambda \\
    &\approx& S(0)\, \big[-\gamma - \ln (\lambda_{\min} t)\big]\;,
\end{eqnarray}
which is equivalent to our results in Eqs.~\eqref{eq:KD} and \eqref{eq:DFormFactor}.
We thus conclude that our system provides yet another instance of glassy logarithmic 
relaxation arising from the interplay between parity symmetry and Anderson localisation.

Note that the behaviour $P_S(s) \sim 1/s$ itself is a consequence of the fact that the localisation centers are uniformly distributed in $[\xi, M]$: indeed, this implies that the distribution of $\ln s$ is uniform over $[\ln s_{\min},\ln s_{\max}]$, which in turn gives the $1/s$ 
dependence of the spacing distribution. This is reminiscent of Benford's law \cite{benford1938law}: aggregating heterogeneous data spanning various orders of magnitude gives numbers whose logarithm is uniformly distributed, which leads to a law 
$\sim 1/d$ for the first digit $d$. Ultimately, the behaviour described in this section has its origin in the fact that there is no privileged scale for hybridization of wavefunctions in momentum space.

\subsection{Crossover time $t_{\max}$}

Roughly speaking, the dynamics starts resolving the doublets when $s_{\max}\tau \gtrsim 1$, 
that is when $t \gtrsim t_{\max} = eN/(2\pi) \approx 0.43 \, N$. 
For the system size $N=32769$ chosen in Fig.~\ref{fig:n+time}, 
we find $t_{\max} \approx 14177$, which is compatible with the data in Fig.~\ref{fig:n+time}b. 

More precisely, the crossover time $t_{\max}$ where the logarithmic dynamics takes over the early-time dynamics can also be estimated by using the asymptotics $M(\tau\gg 1) \approx 1-1/\tau$ of Eq.~\eqref{eq:MickFunc} to describe the monotonically increasing part of $C_B$ and the logarithmically decreasing behaviour to describe the monotonically decreasing part of $C_B$. We identify $t_{\max}$ as the time where the time derivatives of the ascending and descending parts cancel each other. As such, $t_{max}$ should solve $K^\infty_{B}t_H/t^2 = 1/(tN_L)$. Since $t_H=\xi$, we get the estimates $\tau_{\max} = t_{\max}/N= K^\infty_{B}/2$ and $C_{\max}=C_B(x_0,t_{\max}) \sim K^\infty_{B}-2\xi/N$ by identification with $K^\infty_{B} \,M(t_{\max}/t_H)$ and using the asymptotics of $M(x)$. These predictions are in good agreement with our numerical data shown in panels (e) and (f) of Fig.~\ref{fig:n+time}.

In Appendix B, we discuss some features of the intermediate time dynamics around $t_{\max}$ that are relevant to Fig.~\ref{fig:3 regimes} in the localised and metallic regimes. 

\section{Conclusion}

In this paper, we have revisited the localisation dynamics of a paradigmatic model 
of quantum chaos and Anderson localisation, the quantum kicked rotor \cite{chirikov1979universal,Casati1979,haake1991quantum}, 
focusing on a regime relevant to recent cold-atom experiments 
\cite{arrouas2025coherent}: the case where the pseudo-disorder possesses parity symmetry. 
By analysing the coherent scattering interference signatures of localisation, 
namely the coherent backscattering and coherent forward scattering peaks, 
we have uncovered a striking dynamical consequence of this symmetry: 
a logarithmic relaxation reminiscent of that observed in glassy systems driven out of equilibrium \cite{bouchaud1992, bouchaud1998out, amir2012relaxations, PhysRevLett.92.066801, pollak2013electron, PhysRevB.88.085106, PhysRevLett.88.076101}

We first analysed the implications of parity symmetry in the localised regime. 
This symmetry induces doublets of quasi-degenerate states formed by symmetric and antisymmetric 
combinations of exponentially localised states centred at opposite momenta. 
These doublets generate specific spectral correlations that directly control the slow dynamics. 
In particular, the time dependence of the coherent scattering peaks can be related to the spectral form factor and, ultimately, to the nearest-level spacing distribution.

A central result of this work is the emergence of a Shnirelman peak in the level-spacing 
distribution at small spacings $s$, characterized by the power-law behaviour 
$P_S(s) \sim 1/s$ \cite{chirikov1995shnirelman}, which reflects the presence of quasi-degenerate doublets. 
We have shown that this singular contribution is responsible for the logarithmic relaxation. 
We derived analytically and confirmed numerically the associated scaling laws, 
which in our case take the form of a scaling in $t/t_D$, where 
$t_D = N \exp[N/(2\xi)]$ is the characteristic time set by the most weakly split (most distant) Floquet eigenstate parity doublets. 
This time scale plays a role analogous to the waiting time $t_w$ in glassy systems prepared out of equilibrium. 
The resulting scaling behaviour is reminiscent of aging, in the sense that the relaxation depends explicitly on $t_D \equiv t_w$. Interestingly enough, this behaviour also implies that the infinite-size and infinite-time limits do not commute in the localised regime. In the infinite-size limit, doublets proliferate and the dynamics of the system is never able to resolve them all. This means that the system ``freezes" its CFS and CBS contrasts to their maximum values $3$ and $1$ after the localisation time, as essentially observed in Fig.\ref{fig:3 regimes}a.

Despite the strong analogy with glassy dynamics, important differences must be emphasized. 
First, our experimental protocol is fundamentally different: it is a quench and no external waiting time is imposed. 
The scale $t_D$ emerges intrinsically from the interplay between parity symmetry and Anderson localisation in a finite system. 
Second, the evolution is fully quantum and unitary; the relaxation does not originate from coupling to a thermal bath or from decoherence effects. However it does result from quantum tunnelling leading to hybridization.

Our results demonstrate that coherent forward scattering provides a particularly sensitive probe of hidden symmetry properties in disordered quantum systems, as recently suggested experimentally \cite{arrouas2025coherent}. 
More broadly, they reveal a new and robust analogy between Anderson localisation and glassy physics, complementing previously discussed connections in other localisation regimes \cite{PhysRevLett.122.030401, izem2025kardar}.

\begin{acknowledgements}
We dedicate this paper to the memory of the late Dominique Delande, who introduced some of us to the fields of quantum chaos and localisation, in particular in the context of cold atoms, and whose legacy on these topics is invaluable.
    We acknowledge the use of computational resources
at the Calcul en Midi-Pyrénées (CALMIP) for our numerical simulations. This work is supported by the ANR projects QUTISYM (ANR-23-PETQ-0002) and ManyBodyNet, the EUR Grant NanoX No. ANR-17-EURE-0009, the Singapore Ministry of Education Academic Research Funds
Tier II (MOE-T2EP50223-0009 and MOE- T2EP50222-0005), the Région Occitanie, and the Institut Universitaire de France.

\end{acknowledgements}

\emph{{Data availability.}\textemdash}The data that support the findings of
this article are openly available. 

%\bibliographystyle{apsrev4-2}
%\bibliography{REFs.bib}% Produces the bibliography via BibTeX.

\appendix

\section{CFS and CBS peak shape} 

From Eq.~\eqref{eq:x0density}, we get 
\begin{equation}
   n_\infty(x) = n(x, t \to\infty) = \overline{\sum_{a\in\mathcal{H}} \varphi^2_a(x)\varphi^2_a(x_0)}.
   \label{eq:n_infty}
\end{equation}
Since $n_\infty(x)$ is even in $x$, the CBS and CFS peaks at infinite times are twin images of each other. Each is made of bell-shaped pedestal centered at $\pm x_0$ with a spike on top. It is the hybridized nature of the eigenfunctions in momentum space that is responsible for this behaviour. Indeed, writing the even/odd eigenstates $\varphi^\pm_a(p) \sim F_a(p-p_a) \pm F_a(-p+p_a)$ ($p_a=a\hbar_e$ being the localisation center) we have $\varphi^\pm_a(x) \sim (e^{-ip_ax}\pm e^{ip_ax})\,F_a(x)$ where $F_a(x)$ is the Fourier transform of $F_a(p)$. Thus, $|\varphi^\pm_a(x)|^2 \sim 2[1\pm\cos(2p_ax)]\, |F_a(x)|^2$. Eq.~\eqref{eq:n_infty} then involves $\cos[p_a(x\pm x_0)]$ terms which, under disorder averaging, will produce $\delta$-peaks at $\pm x_0$, while the other remaining terms will produce a smooth contribution whose width is given by the inverse localisation length $\xi^{-1}$. The spikes observed in the spatial distribution are thus a direct consequence of the hybridization of wave functions in momentum space and a direct signature in position space of the parity symmetry of the system.\\ 

\section{Intermediate times dynamics in the localised and metallic regimes}

Coming back to Eq.~\eqref{eq:Cminus} for times $t_H \lesssim t \ll t_D$ in the localised regime $\xi \ll M$,  
one can set $\exp(-i\omega_{ab}t) \approx 0$ for the opposite-parity levels which are resolved after $t$ and set $\exp(-i\omega_{ab}t) \approx 1$ for the set $D(a,b)$ of unresolved doublets levels. We expect their number to be essentially given by $N_S$. For $t_H \lesssim t \ll t_D$, we thus have:
\begin{equation}
    C_B(x_0,t) \approx 2N \overline{\sum_{D(a,b)} \varphi^2_a(x_0) \, \varphi^2_b(x_0)} \approx
    C_{\max}
    \label{eq:overshoot}
\end{equation}
and we get 
\begin{eqnarray}
\label{cinfx0}
    &&C(x_0,t) \approx C_\infty+C_{\max} 
    \\
    &&C(-x_0,t) \approx C_\infty -C_{\max} 
\end{eqnarray}
at times $t_H \lesssim t \ll t_D$. This also shows that, at intermediate times, $C_\infty = [C(x_0,t)+C(-x_0,t)]/2$ and $C_{\max}= [C(x_0,t)-C(-x_0,t)]/2$. Neglecting correlations between eigenstates in Eq.~\eqref{eq:overshoot} and assuming uniform spread in position space, we get the estimate $C_{\max} \sim 2 \sigma_S$, a result consistent with the behaviour obtained from Eq.~\eqref{eq:KD} at $s_{\max}\tau \ll 1$. We thus infer
that, at intermediate times, the CFS contrast overshoots the value $C_\infty$ by an amount scaling like $2\sigma_S$ while the CBS contrast undershoots $C_\infty$ by the same amount. We also infer that $K^\infty_B \sim 1$ and $t_{\max} \sim N/2$, compatible with the first estimate $0.43\,N$.

For comparison, we have $C_{\max} \sim 0$ in the metallic regime ($N_L \lesssim 1$) and we get $C(\pm x_0,t) \to 2$. This is what is observed in Fig.~\ref{fig:3 regimes}b,c. In the strongly localised regime ($N_L \gg 1$), we need to consider 2 cases. When $\xi/\ell \sim K/\hbar_e\gg 1$, then $C_\infty =2$ and $C_{\max}\sim 1$, such that $C(x_0,t) = 3$ and $C(-x_0,t)= 1$. This behaviour is somewhat reminiscent of the symplectic case where the spectral form factor overshoots its infinite-time limit unit value \cite{akkermans2007, liu2018spectral}, the doublets posturing effective spin states. However, when $\xi/\ell \sim 1$, which happens in Fig.~\ref{fig:3 regimes}a, it is known that $C_\infty$ is reduced by a quantity scaling like $(\ell/\xi)^{d/2}$ ($d$ is the system dimension) through a central limit theorem (CLT) argument \cite{lee2014dynamics, stein1972bound}. Since $C_{\max}$ and $C_\infty$ involve the same type of eigenstates correlations, we expect $C_{\max}$ to be also reduced by the same quantity. This mean that the CBS contrast should remain immune to these CLT effects as they would cancel each other, while the CFS contrast should be reduced as these corrections would add up. As a consequence, the CBS contrast should stick to $C(-x_0,t)=1$ at intermediate times, as is observed in Fig.~\ref{fig:3 regimes}a, while the CFS contrast should be reduced from its nominal value $C(x_0,t) =3$ at intermediate times. From Fig.~\ref{fig:3 regimes}a, we get $C_\infty \sim 1.75$ and $C_{\max} \sim 0.75$, implying a CLT-correction of about $0.25$, which is of the same order of magnitude as $\sqrt{\ell/\xi} \sim \sqrt{\hbar_e/K} = 0.4$. To go further, a more careful analysis of the wavefunction correlations when $\xi \sim \ell$ is needed. We leave this for future work.
\\

\end{document}